\documentclass{aa}
\usepackage{graphicx}
\usepackage{color}
\usepackage{natbib}
\usepackage{savefnmark}
\bibpunct{(}{)}{;}{a}{}{,}
\usepackage{txfonts}
\begin{document} 

\def\sun{\hbox{$\odot$}}
\def\degr{\hbox{$^\circ$}}
\def\arcmin{\hbox{$^\prime$}}
\def\arcsec{\hbox{$^{\prime\prime}$}}

   \title{Major impact from a minor merger}

   \subtitle{The extraordinary hot molecular gas flow in the Eye of the NGC~4194 Medusa galaxy \thanks{Based on observations carried out under project 
    	     number I15AA001 with the IRAM NOEMA Interferometer. IRAM is supported by INSU/CNRS (France), MPG (Germany) and IGN (Spain).}}

   \author{S. K\"onig
          \inst{1,2}
          \and
          S. Aalto
	   \inst{1}
          \and
          S. Muller
	   \inst{1}
	  \and
          J.~S. Gallagher III
	   \inst{3}
	  \and
          R.~J. Beswick
	   \inst{4}
	  \and
          E. Varenius
	   \inst{1,4}
	  \and
          E. J\"utte
	   \inst{5}
         \and
          M. Krips
          \inst{2}
         \and
          A. Adamo
          \inst{6}
          }

   \institute{Chalmers University of Technology, Department of Space, Earth and Environment, Onsala Space Observatory, 43992 
	     Onsala, Sweden\\
              \email{sabine.koenig@chalmers.se}
         \and
             Institut de Radioastronomie Millim\'etrique, 300 rue de la Piscine, Domaine Universitaire, F-38406 Saint 
  	     Martin d'H\`eres, France
         \and
             Department of Astronomy, University of Wisconsin, 475 N. Charter Street, Madison, WI, 53706, USA
         \and
             University of Manchester, Jodrell Bank Centre for Astrophysics, Oxford Road, Manchester, M13 9PL, UK
	 \and 
	     Astronomisches Institut Ruhr-Universit\"at Bochum, Universit\"atsstra\ss e 150, 44780 Bochum, Germany
	 \and 
	     Department of Astronomy, Oskar Klein Centre, Stockholm University AlbaNova University Centre, 106\,91 Stockholm, Sweden
             }

   \date{Received ; accepted }

  \abstract
   {Minor mergers are important processes contributing significantly to how galaxies evolve across the age of the Universe. Their impact on the growth 
    of supermassive black holes (SMBHs) and star formation is profound -- about half of the star formation activity in the local Universe is the result 
    of minor mergers.}
   {The detailed study of dense molecular gas in galaxies provides an important test of the validity of the relation between star formation rate (SFR) 
    and HCN luminosity on different galactic scales -- from whole galaxies to giant molecular clouds in their molecular gas-rich centers.}
   {We use observations of HCN and HCO$^{\rm +}$\,1$-$0 with NOEMA and of CO\,3$-$2 with the SMA to study the properties of the dense molecular gas 
    in the Medusa merger (NGC~4194) at 1\arcsec\ resolution. In particular, we compare the distribution of these dense gas tracers with CO\,2$-$1 
    high-resolution maps in the Medusa merger. To characterise gas properties, we calculate the brightness temperature ratios between the three tracers 
    and use them in conjunction with a non-LTE radiative line transfer model.}
   {The gas represented by HCN and HCO$^{\rm +}$\,1$-$0, and CO\,3$-$2 does not occupy the same structures as the less dense gas associated with the 
    lower-$J$ CO emission. Interestingly, the only emission from dense gas is detected in a 200~pc region within the ``Eye of the Medusa'', an asymmetric 
    500~pc off-nuclear concentration of molecular gas. Surprisingly, no HCN or HCO$^{\rm +}$ is detected for the extended starburst of the Medusa merger. 
    Additionally, there is only little HCN or HCO$^{\rm +}$ associated with the AGN. The CO\,3$-$2/2$-$1 brightness temperature ratio inside ``the Eye'' 
    is $\sim$2.5 -- the highest ratio found so far, and implying optically thin CO emission. The CO\,2$-$1/HCN\,1$-$0 ($\sim$9.8) and 
    CO\,2$-$1/HCO$^{\rm +}$\,1$-$0 ($\sim$7.9) ratios show that the dense gas filling factor must be relatively high in the central region, consistent 
    with the elevated CO\,3$-$1/2$-$1 ratio.}
   {The line ratios reveal an extreme, fragmented molecular cloud population inside ``the Eye'' with large bulk temperatures (T\,$>$\,300~K) and high gas 
    densities (n(H$_{\rm 2}$)\,$>$\,10$^{\rm 4}$~cm$^{\rm -3}$). This is very different from the cool, self-gravitating structures of giant molecular 
    clouds normally found in the disks of galaxies. ``The Eye of the Medusa'' is found at an interface between a large-scale minor axis inflow and the 
    central region of the Medusa. Hence, the extreme conditions inside ``the Eye'' may be the result of the radiative and mechanical feedback from a 
    deeply embedded, young and massive super star cluster, formed due to the gas pile-up at the intersection. Alternatively, shocks from the inflowing 
    gas entering the central region of the Medusa may be strong enough to shock and fragment the gas. For both scenarios, however, it appears that the 
    HCN and HCO$^{\rm +}$ dense gas tracers are not probing star formation, but instead a post-starburst and/or shocked ISM that is too hot and 
    fragmented to form new stars. Thus, caution is advised in linking the detection of emission from dense gas tracers to evidence of ongoing or imminent 
    star formation.}

   \keywords{galaxies: evolution -- 
		galaxies: individual: NGC~4194 -- 
		galaxies: starburst -- 
		galaxies: active -- 
		radio lines: ISM -- 
		ISM: molecules
               }

\titlerunning{Dense gas in the ``Eye of the Medusa''}

   \maketitle
%

\section{Introduction} \label{sec:intro}

\begin{figure*}[h]
 \centering
  \begin{minipage}[hbt]{0.33\textwidth}
  \centering
    \includegraphics[width=0.975\textwidth]{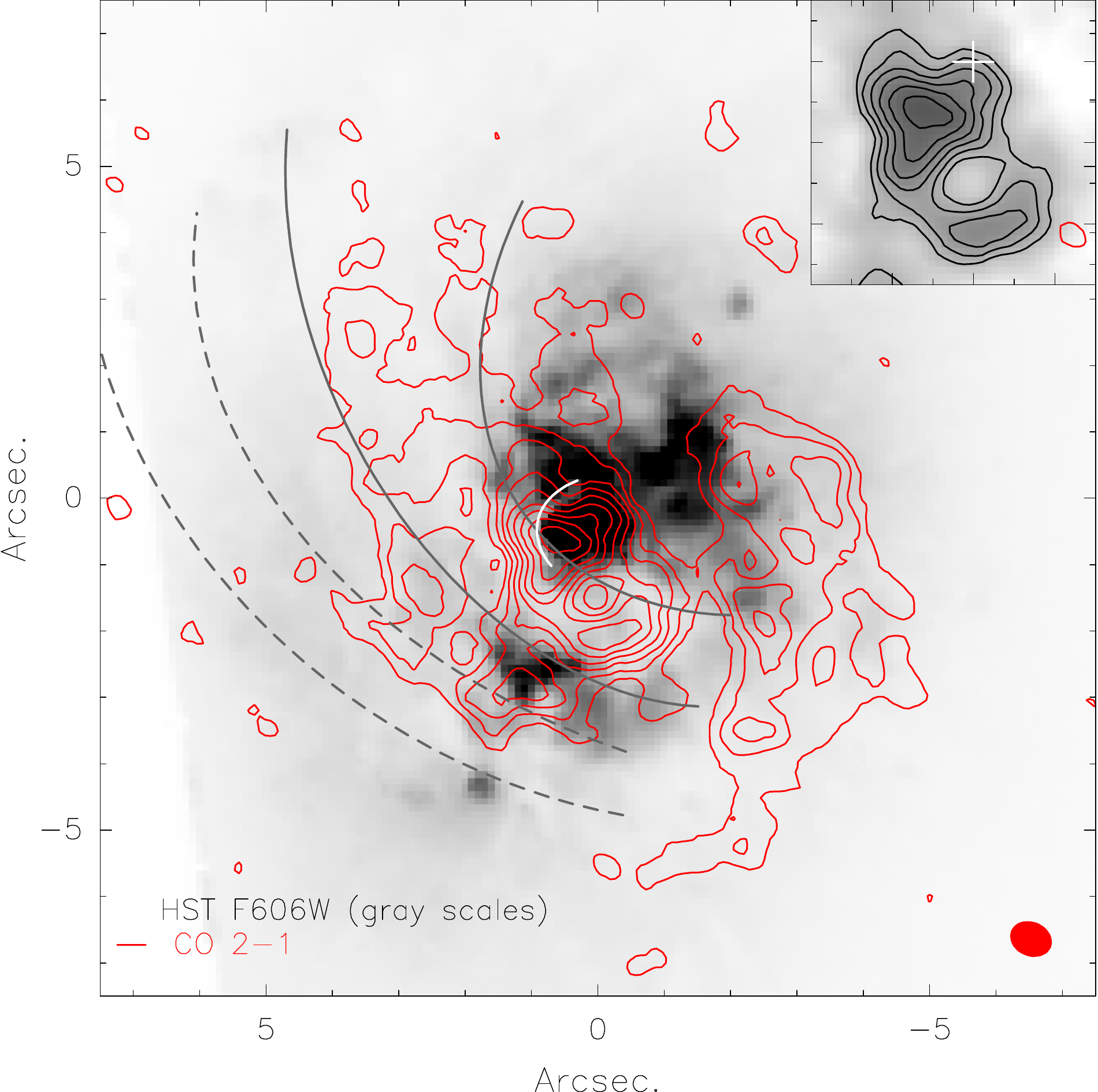}
  \end{minipage}
  \begin{minipage}[hbt]{0.33\textwidth}
  \centering
    \includegraphics[width=0.975\textwidth]{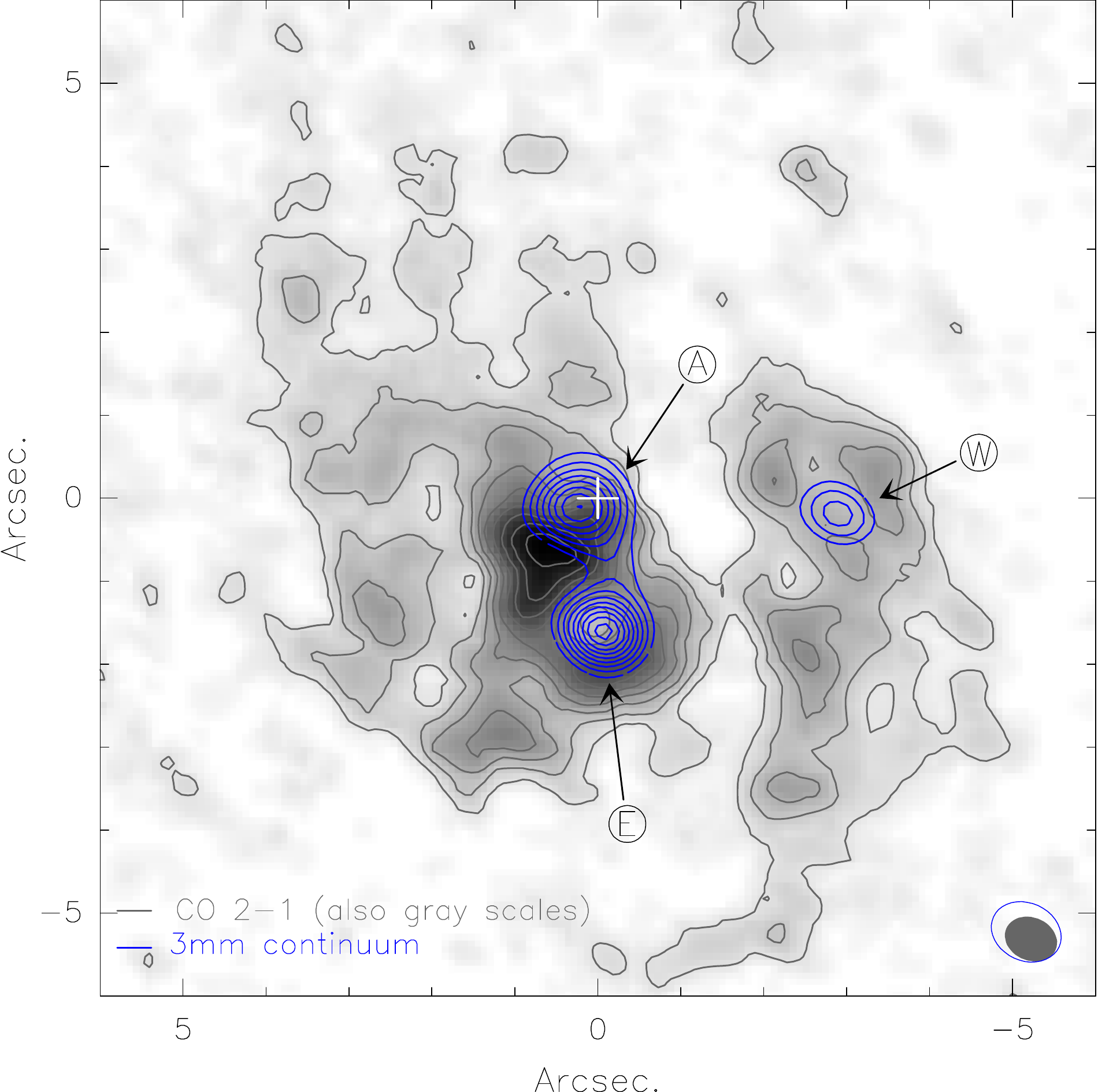}
  \end{minipage}
  \begin{minipage}[hbt]{0.33\textwidth}
  \centering
    \includegraphics[width=0.975\textwidth]{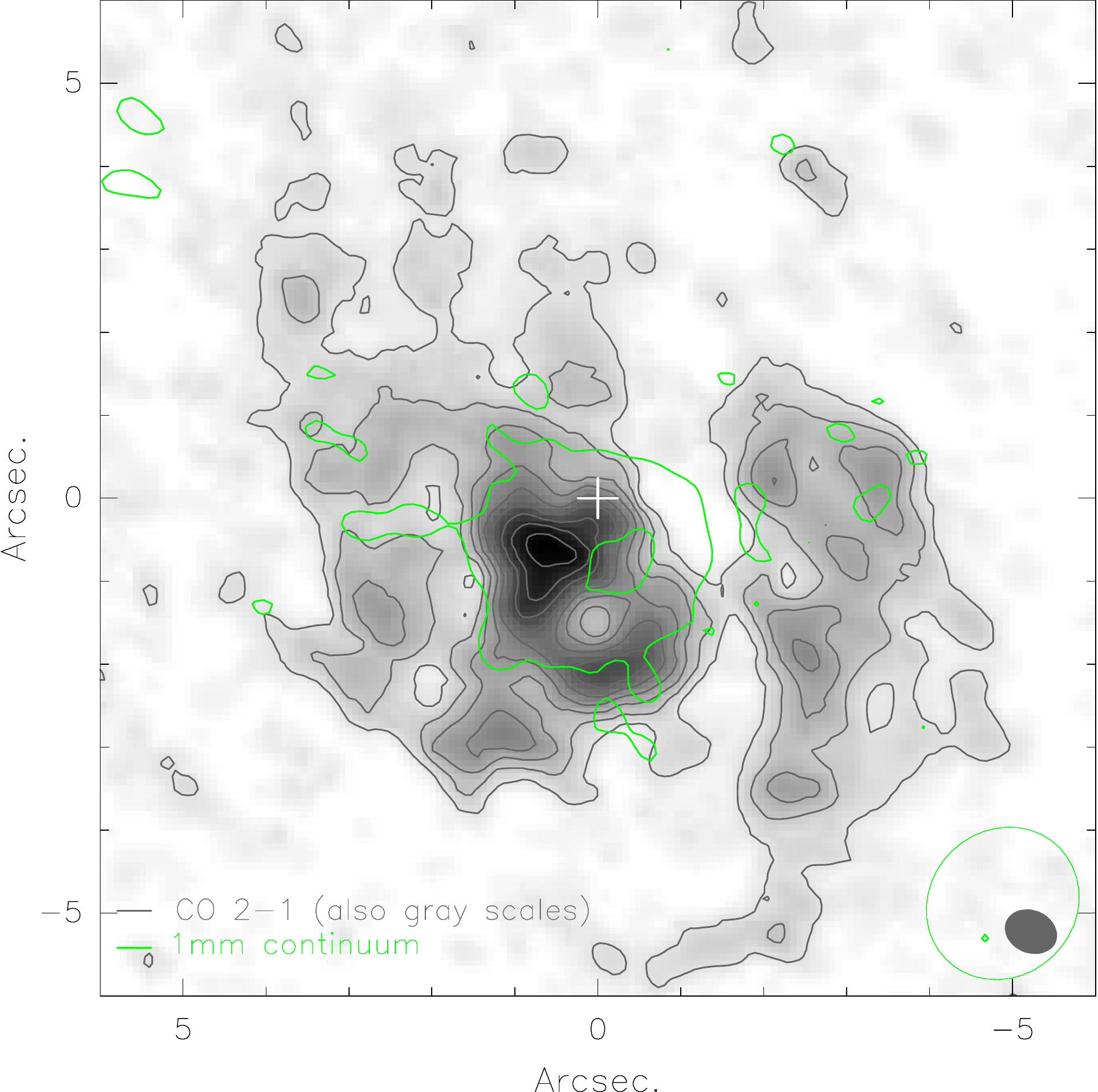}
  \end{minipage}
  \caption{\footnotesize \textit{Left:} Overlay of the high resolution CO\,2$-$1 emission contours on top of an HST WFPC2 \textit{F606W} filter image 
   \citep[from][]{koenig14}. The insert shows a zoom into the 3\arcsec\,$\times$\,3\arcsec\ surrounding the ``Eye of the Medusa''. Contours start 
   at 2$\sigma$ and are spaced in steps of 2$\sigma$ (1$\sigma$ $\sim$0.7~Jy\,beam$^{\rm -1}$\,km\,s$^{\rm -1}$). The locations of the most important 
   dust lanes are indicated by grey and white curves. \textit{Center:} Distribution of the integrated intensity emission of the 3~mm continuum (blue 
   contours) compared to the CO\,2$-$1 emission (grey contours and background). The 3~mm continuum contours start at 5$\sigma$ and are spaced in steps of 
   3$\sigma$ (1$\sigma$ $\sim$50~$\mu$Jy\,beam$^{\rm -1}$). The main 3~mm continuum emission peak is located at the center of the ``Eye of the 
   Medusa'' (E). \textit{Right:} 1~mm continuum emission on top of the CO\,2$-$1 emission. Contours are at 3 and 6$\sigma$ (1$\sigma$ 
   $\sim$0.5~mJy\,beam$^{\rm -1}$). The emission peaks south of the AGN position, close to the center of ``the Eye''. North is up, east to the left. The 
   position of the 1.4~GHz continuum peak is marked by a white cross \citep{bes05}. Beam sizes are 0.6\arcsec\,$\times$\,0.5\arcsec\ for CO\,2$-$1, 
   0.9\arcsec\,$\times$\,0.7\arcsec\ for the 3~mm continuum, and 1.9\arcsec\,$\times$\,1.8\arcsec\ for the 1~mm continuum.}
  \label{fig:co2-1+3mmcont}
\end{figure*}

Galaxy evolution is a fundamental part of the overall evolution of the Universe. Its sphere of influence extends from the large scales dominated by 
dark matter, down to the small scales ruled by dissipative baryons that can form stars and grow supermassive black holes \citep[SMBHs,][]{shl13}. 
Interactions and mergers are a known and efficient mechanism for galaxy growth \citep[e.g.,][]{sandage90,sandage05,kor04}. Minor mergers (unequal mass 
progenitors, mass ratios: $\gtrsim$1:4) occur much more frequently than major mergers \citep[equal mass progenitors, e.g.,][]{her95,kav09,lot11}. Their 
impact on the growth of SMBHs and star formation is profound -- about half of the star formation activity in the local Universe is the result of minor 
mergers \citep{kav14,kav16}.\\ 
\indent
\citet{schm59} was the first to systematically study the connection between gas density and star formation rate (SFR) in the Milky Way. Using 
H$\alpha$, CO\,1$-$0 and HI, \citet{ken98} determined star formation rate surface densities and gas surface densities in a sample of normal 
spirals and starburst galaxies - confirming the results of \citet{schm59}. \citet{gao04b,gao04a} found a similar, but tighter correlation between 
gas density and SFR when using HCN to study the properties of dense gas in relation to star formation in luminous infrared galaxies (LIRGs), 
ultra-luminous infrared galaxies (ULIRGs) and normal spiral galaxies.\\
\indent
The \object{Medusa merger} \citep[\object{NGC~4194}, D\,=\,39~Mpc, 1\arcsec\,=\,189~pc][]{bes05} is a minor merger that harbours a region of highly 
efficient star formation in its inner 2~kpc -- the star formation efficiency \citep[SFE, $\sim$1.5\,$\times$\,10$^{\rm -8}$~yr$^{\rm -1}$,][]{koenig14} 
rivals even that of well-known ULIRGs \citep{aalto00}. Low- to intermediate-density gas spans the main body of NGC~4194 -- from the tidal tail at 4.7~kpc 
radius, down to the central starburst \citep{aalto01}. High-resolution CO observations have shown that molecular gas is associated with the minor axis 
dust lane crossing the galaxy's main body \citep{aalto00}. The brightest CO emission is found in a striking, off-nuclear structure called the ``Eye of 
the Medusa'' \citep{koenig14}. A fraction of the star formation in NGC~4194 is going on in young super star clusters 
\citep[SSCs, 5-15~Myr,][]{bon99,pel07} with a kpc-scale distribution \citep{wei04,han06}.\\
\indent
Single-dish observations revealed a global CO-to-HCN\,1$-$0 luminosity ratio of $\sim$25 for NGC~4194 within a 29\arcsec\ beam \citep{costa11}, a 
significantly higher value than for ULIRGs \citep[$\sim$6, e.g.,][]{sol92,cur00}, indicating that the average fraction of dense gas is 
significantly lower despite the similar extreme SFE.\\
\indent
Here we present a high-angular resolution study of the dense molecular gas in the Medusa merger, using HCN, HCO$^{\rm +}$ and CO emission as tracers 
thereof. Throughout the paper, we are concerned with pure rotational transitions of HCN, HCO$^{\rm +}$, CO, C$_{\rm 2}$H, SiO, H$^{\rm 13}$CO$^{\rm +}$ 
between upper state $J'$\,=\,$j$ and lower state $J$\,=\,$i$ that are labeled $j$\,$-$\,$i$.\\
\indent
In Sect.\,\ref{sec:obs} we describe the observations and how the data were reduced and analysed, in Sect.\,\ref{sec:results} we present the 
results, and in Sect.\,\ref{sec:discussion} we discuss their implications.


\section{Observations} \label{sec:obs}

\subsection{NOEMA observations} \label{subsec:obs_noema}

The HCN and HCO$^{\rm +}$\,1$-$0 observations of \object{NGC~4194} were carried out with the Northern Extended Millimeter Array (NOEMA) in the frame of 
science verification on March 10, 2015. Data were taken with seven antennas in extended configuration, with baselines between 32 and 760~m. Thus, the 
observations are sensitive to scales smaller than 13.5\arcsec. The phase center of the observations was located at $\alpha$=12:14:09.660 and 
$\delta$=+54:31:35.85 -- the 1.4~GHz radio continuum peak \citep[J2000,][]{bes05}. The 3~mm-band receivers were tuned to 87.039~GHz to cover the HCN and 
HCO$^{\rm +}$\,1$-$0 lines in the 3.6~GHz bandwidth of WideX. Also, the C$_{\rm 2}$H, H$^{\rm 13}$CO$^{\rm +}$\,1$-$0, and SiO\,2$-$1 lines were 
accessible in the same tuning. The instrumental spectral resolution was 1.95~MHz ($\sim$6.7~km\,s$^{\rm -1}$). For analysis purposes we smoothed the 
data to $\sim$10~km\,s$^{\rm -1}$, resulting in 1$\sigma$ rms noise levels per channel of $\sim$0.4~mJy\,beam$^{\rm -1}$ for both HCN and 
HCO$^{\rm +}$\,1$-$0. The antenna system noise temperature $T_{\rm sys}$ ranged from 70 to 160~K. During the observations, different sources were 
observed for calibration purposes: \object{MWC~349} and \object{LkHa~101} as flux calibrators, \object{3C~84} as bandpass calibrator, and 
\object{J1150+497} and \object{J1259+516} as phase calibrators. We estimate flux calibration uncertainty of about 15-20\%.\\
\indent
Data reduction and analysis were performed using the CLIC and MAPPING software packages within GILDAS\footnote{http://www.iram.fr/IRAMFR/GILDAS}. 
Applying a natural weighting scheme led to a nearly circular beam size of $\sim$1.0\arcsec.

\begin{figure*}[ht]
  \begin{minipage}[hbt]{0.33\textwidth}
  \centering
    \includegraphics[width=0.975\textwidth]{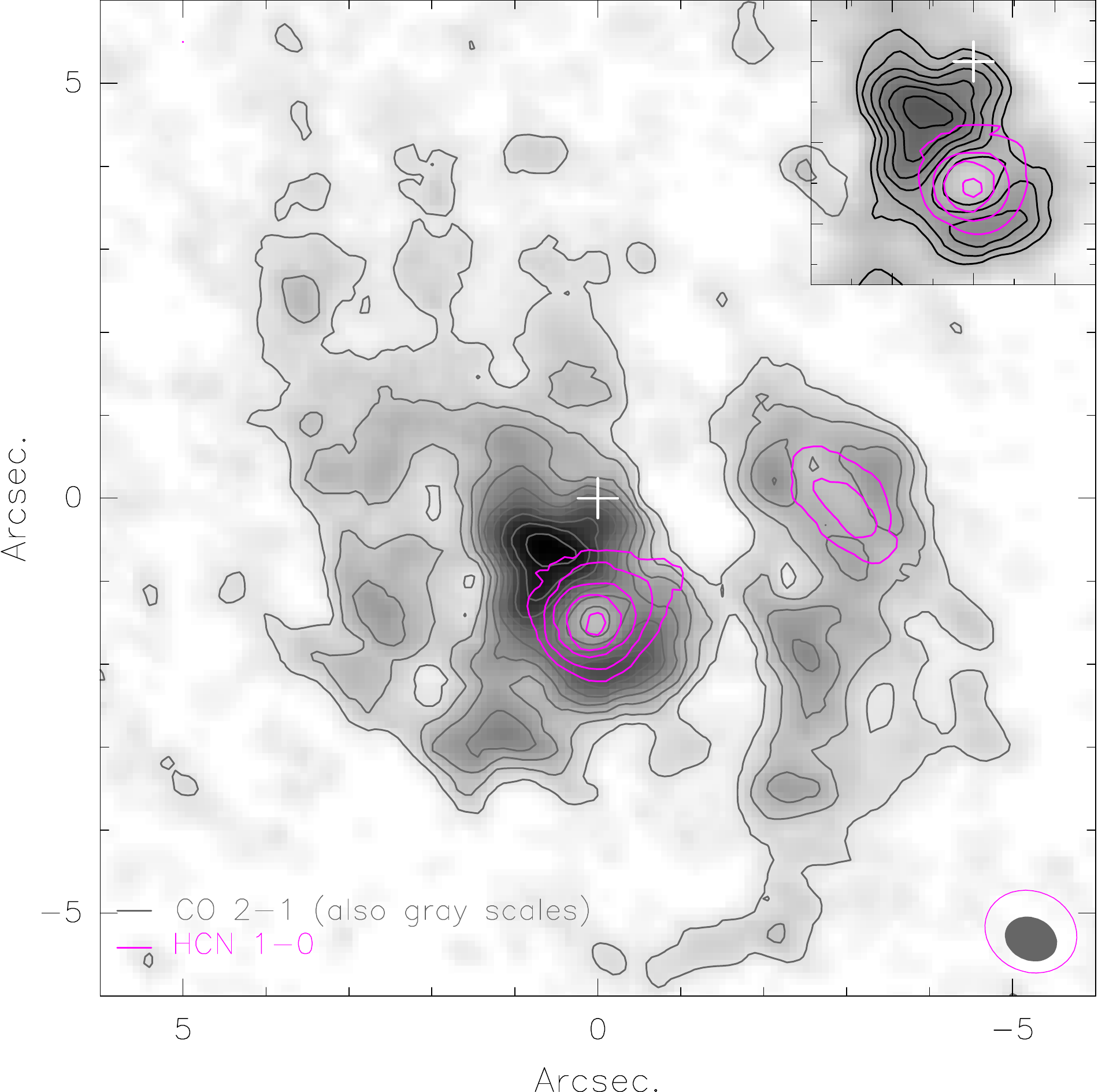}
  \end{minipage}
  \begin{minipage}[hbt]{0.33\textwidth}
  \centering
    \includegraphics[width=0.975\textwidth]{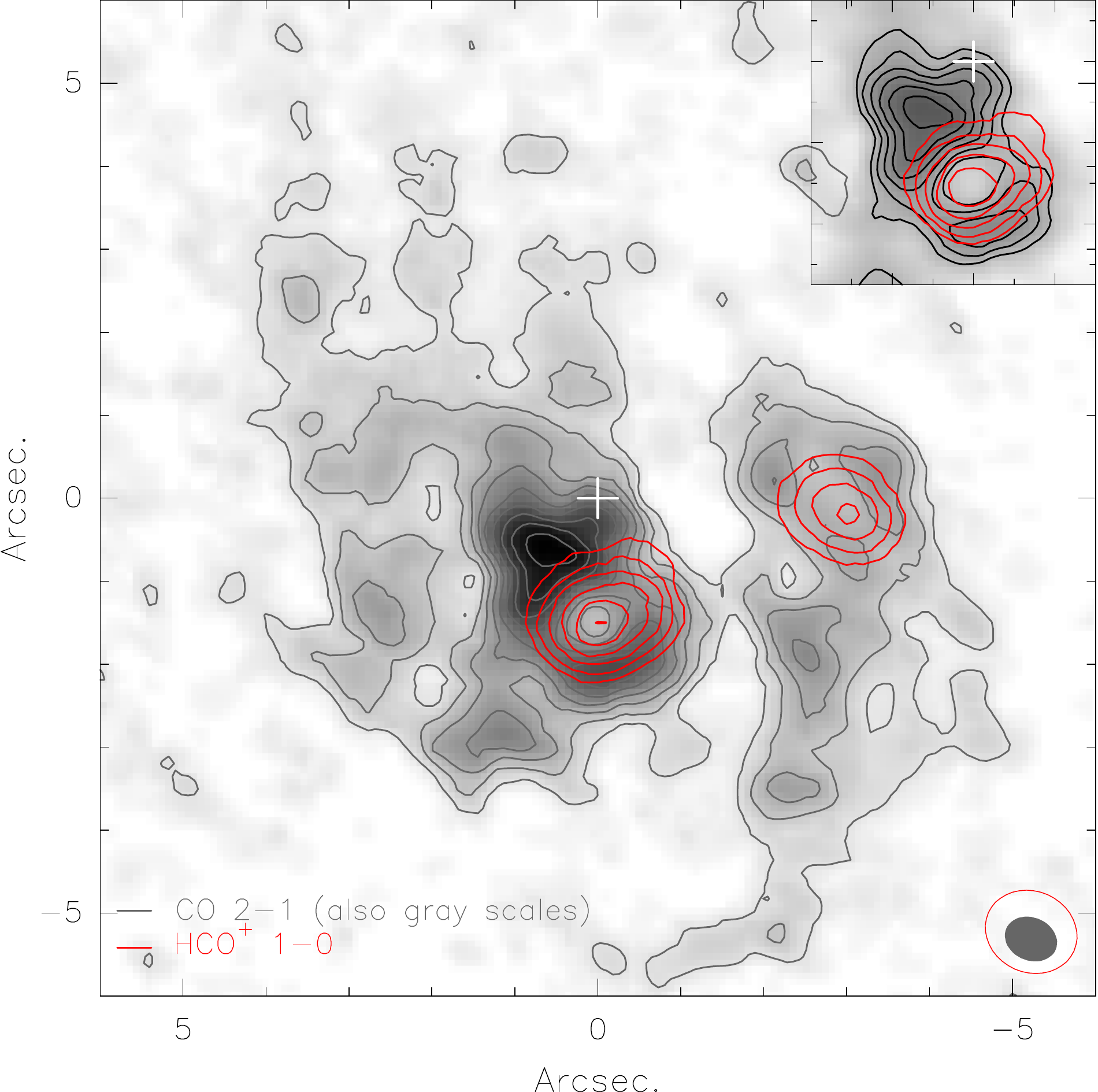}
  \end{minipage}
  \begin{minipage}[hbt]{0.33\textwidth}
  \centering
    \includegraphics[width=0.975\textwidth]{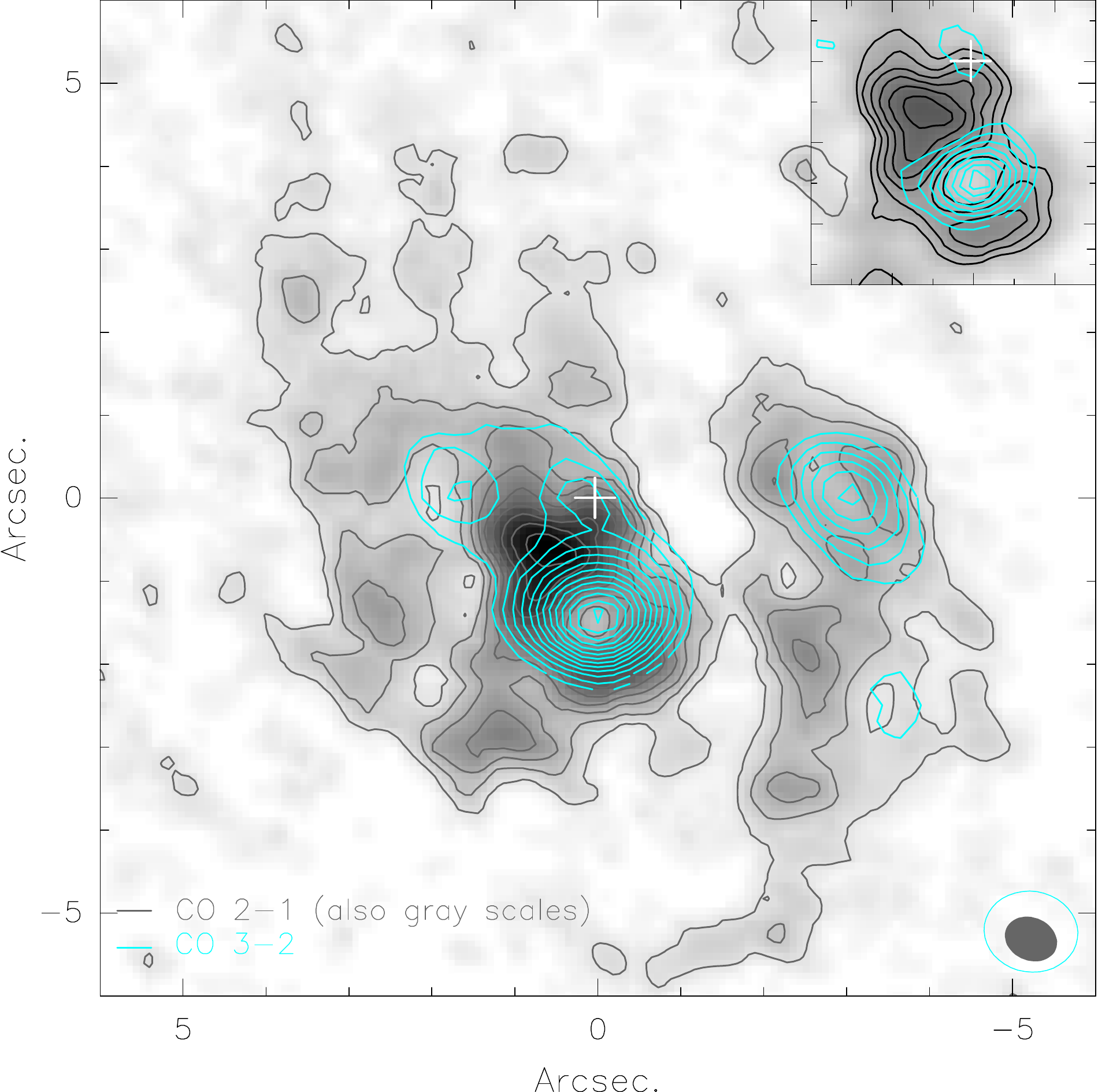}
  \end{minipage}
  \caption{\footnotesize Integrated intensity distributions of the HCN\,1$-$0 (\textit{left}, in magenta), HCO$^{\rm +}$\,1$-$0 (\textit{centre}, in 
   red) and CO\,3$-$2 emission contours (\textit{right}, in light blue) at the same spatial resolution ($\sim$1.1\arcsec\,$\times$\,1.0\arcsec) on top of 
   the CO\,2$-$1 high-resolution emission \citep[background and grey contours,][]{koenig14}. Contours start at 5$\sigma$ and are spaced in steps of 
   3$\sigma$ (1$\sigma$(HCN\,1$-$0) $\sim$20.2~mJy\,beam$^{\rm -1}$\,km\,s$^{\rm -1}$, 1$\sigma$(HCO$^{\rm +}$\,1$-$0) 
   $\sim$21.2~mJy\,beam$^{\rm -1}$\,km\,s$^{\rm -1}$, 1$\sigma$(CO\,3$-$2) $\sim$1.6~Jy\,beam$^{\rm -1}$\,km\,s$^{\rm -1}$). The inset in each of the 
   images shows the higher resolution data for each tracer (beam sizes: 1.0\arcsec\,$\times$\,0.8\arcsec, 1.0\arcsec\,$\times$\,0.8\arcsec, 
   0.7\arcsec\,$\times$\,0.7\arcsec). North is up, east to the left. The position of the 1.4~GHz continuum peak is marked by a white cross 
   \citep{bes05}.}
  \label{fig:hcn1-0+hco+1-0+co3-2}
\end{figure*}

\subsection{SMA observations} \label{subsec:obs_sma}

\subsubsection{CO\,3$-$2} \label{subsec:obs_sma_co3-2}

On April 10, 2016, additional CO\,3$-$2 observations were obtained with the Submillimeter Array (SMA) in its extended configuration with baselines 
between 44 and 226~m. The 345~GHz receivers, tuned to 342.935~GHz, were used in conjunction with the correlator in 4~GHz mode with 128 channels and a 
spectral resolution of $\sim$0.8~MHz ($\sim$0.7~km\,s$^{\rm -1}$). We smoothed the data to a velocity resolution of $\sim$10.6~km\,s$^{\rm -1}$, which 
results in a 1$\sigma$ rms noise level per channel of $\sim$29.9~mJy\,beam$^{\rm -1}$. The observed data set is sensitive to scales smaller than 
$\sim$2.5\arcsec\ at the observing frequency. For calibration, \object{Callisto} and \object{MWC349a} (flux), \object{J1924-292} (bandpass), and 
\object{J1419+543} and \object{J1153+495} (phase) were observed. The SMA data were calibrated using the dedicated MIR/IDL SMA reduction package. The 
estimated flux calibration uncertainty is of the order of 20\%. The visibilities were converted into FITS format and transferred to the GILDAS/MAPPING 
package for further imaging. Similar to the NOEMA observations, the phase center is located at $\alpha$=12:14:09.660 and $\delta$=+54:31:35.85. With 
natural weighting the resulting synthesized beam is $\sim$1.1\arcsec\,$\times$\,1.0\arcsec\ with a position angle of $\sim$81\degr, the beam of the 
uniformly weighted data cube is $\sim$0.74\arcsec\,$\times$\,0.66\arcsec\ (PA $\sim$99\degr).

\subsubsection{1~mm continuum} \label{subsec:obs_sma_1mmcont}

Observations of the $^{\rm 12}$CO and $^{\rm 13}$CO\,2$-$1 line transitions with the SubMillimeter Array (SMA) were taken in compact (April 08, 2010) 
and very extended configuration (February 21, 2010). We extracted the 1~mm continuum information from these data sets. The phase center of the image 
presented in this work is located at $\alpha$=12:14:09.660 and $\delta$=+54:31:35.85. The resulting synthesized beam is 
$\sim$1.9\arcsec\,$\times$\,1.8\arcsec\ (PA $-$46\degr). Imaging the compact configuration alone results in a synthesized beam of 
$\sim$3.4\arcsec\,$\times$\,3.3\arcsec, PA 22\degr. Baseline lengths range between 38 and 509~m. Thus these data are sensitive to emission 
originating from scales smaller than $\sim$4.2\arcsec\ at the observing frequency of 230~GHz. We estimated calibration uncertainties on all fluxes 
of $\sim$20\%. For more details regarding data reduction, array combination and analysis, see \citet{koenig14}.

\begin{table*}[!ht]
\begin{minipage}[!h]{\textwidth}
\centering
\renewcommand{\footnoterule}{}
\caption{\small Characterising properties of the interferometric observations.}
\label{tab:interfero_info}
\tabcolsep0.1cm
\begin{tabular}{lcccccc}
\hline
\noalign{\smallskip}				
\hline
\noalign{\smallskip}
Line                 & $\nu$$_{\rm obs}$ & Observatory & Configuration  & B$_{\rm min}$\footnote{Shortest baseline in the data set.} & B$_{\rm max}$\footnote{Longest baseline in the data set.} & $\Theta$$_{\rm MRS}$\footnote{Maximum recoverable scale. The interferometer in the given configuration is sensitive to structures smaller than this size scale.}\\
\noalign{\smallskip}
                     & [GHz]            &             &                & [m]           & [m]           & [arcsec] \\
\noalign{\smallskip}
\hline
\noalign{\smallskip}
CO\,3$-$2            & 342.935 & SMA   & Extended                & 44 & 226 & 4.1  \\
CO\,2$-$1            & 228.631 & SMA   & Compact + Very Extended & 38 & 509 & 7.1  \\
HCO$^{\rm +}$\,1$-$0 & 88.451  & NOEMA & 7A-Special              & 32 & 760 & 21.8 \\
HCN\,1$-$0           & 87.899  & NOEMA & 7A-Special              & 32 & 760 & 22.0 \\
\noalign{\smallskip}
\hline
\end{tabular}
\end{minipage}
\end{table*}


\section{Results} \label{sec:results}

\subsection{Continuum} \label{subsec:cont}

\subsubsection{3~mm continuum} \label{subsubsec:3mmcont}

The 3~mm continuum, as pictured in Fig.\,\ref{fig:co2-1+3mmcont}, shows three distinct, point-like sources in the center of NGC~4194 -- one at the 
position of the AGN (A), one inside the ``Eye'' (E), and one in the western arm (W) of the CO\,2$-$1 distribution \citep{koenig14}. The highest peak 
flux is found in the emission peak inside the ``Eye'' (see Fig.\,\ref{fig:co2-1+3mmcont}). The flux density recovered from the three components is 
5.6\,$\pm$\,0.4~mJy (within 3$\sigma$ contours): 2.2\,$\pm$\,0.4~mJy for the AGN position (A), 2.0\,$\pm$\,0.4~mJy inside the ``Eye'' (E), and 
1.2\,$\pm$\,0.2~mJy in the western arm (W).

\subsubsection{1~mm continuum} \label{subsubsec:1mmcont}

Fig.\,\ref{fig:co2-1+3mmcont} shows the 1~mm continuum distribution in the central $\sim$2~kpc of NGC~4194. The emission peak is associated with the 
``Eye of the Medusa'', slightly south of the AGN position. The integrated intensity within 3$\sigma$ is 4.1$\pm$0.5~mJy at a central frequency of 
230.5~GHz, when combining compact and very extended configuration. Imaging only the compact configuration yields an integrated intensity of 
15.5$\pm$0.5~mJy. In contrast to the 3~mm continuum emission, the peak associated with ``the Eye'' is of more circular shape 
(Figs.\,\ref{fig:co2-1+3mmcont}, \ref{fig:app_cont}). Also, whereas the 3~mm continuum shows a secondary emission peak in the western arm, the 1~mm 
continuum does not. For a side-by-side comparison of the 3~mm and 1~mm continuum emission distributions degraded to the spatial resolution of the 1~mm 
observations see Fig.\,\ref{fig:app_cont}.


\subsection{Line emission} \label{subsec:lines}

\subsubsection{HCN \& HCO$^{\rm +}$\,1$-$0} \label{subsubsec:hcn_hco+1-0}

The distributions of the HCN and HCO$^{\rm +}$\,1$-$0 emission are quite similar to each other in NGC~4194 (Fig.\,\ref{fig:hcn1-0+hco+1-0+co3-2}): the 
main emission peak is located inside the ``Eye of the Medusa'' (``E'') where CO\,2$-$1 emission shows a distinct minimum 
\citep[Fig.\,\ref{fig:co2-1+3mmcont},][]{koenig14}. This location is also situated inside the minor axis dust lane crossing the main body of NGC~4194. 
A secondary peak is located in the western arm of the CO\,2$-$1 distribution (``W''). A close look at the HCO$^{\rm +}$ channel map also 
reveals low-level emission (at a 3$\sigma$ level) associated with the AGN position (``A'').\\
\indent
The overall HCO$^{\rm +}$\,1$-$0 emission line is brighter than the corresponding line transition of HCN (cf. Fig.\,\ref{fig:3mm_spectrum}). Measuring 
the intensities in the naturally weighted map with 10~km\,s$^{\rm -1}$ wide channels results in integrated fluxes of 
1.0\,$\pm$\,0.1~Jy\,km\,s$^{\rm -1}$ for the HCO$^{\rm +}$ (654\,$\pm$\,61~mJy\,km\,s$^{\rm -1}$ in the ``Eye'', 374\,$\pm$\,48~mJy\,km\,s$^{\rm -1}$ in 
the western arm) and 0.8\,$\pm$\,0.1~Jy\,km\,s$^{\rm -1}$ for HCN (515\,$\pm$\,53~mJy\,km\,s$^{\rm -1}$ in the ``Eye'', 
299\,$\pm$\,44~mJy\,km\,s$^{\rm -1}$ in the western arm), respectively.


\subsubsection{CO\,3$-$2} \label{subsubsec:co3-2}

Like HCN and HCO$^{\rm +}$\,1$-$0, CO\,3$-$2 has its main peak inside the ``Eye'' (``E''). Other emission regions are located in the western arm (``W''), 
at the AGN position (``A'') and east of the AGN in the dust lane. Except for the vicinity of the AGN, these regions are all deficient in CO\,2$-$1 
(Fig.\,\ref{fig:hcn1-0+hco+1-0+co3-2}). The total integrated intensity of all components combined, in the lower resolution cube, amounts to 
158.0\,$\pm$\,7.7~Jy\,km\,s$^{\rm -1}$ (93.8\,$\pm$\,4.9~Jy\,km\,s$^{\rm -1}$ in ``the Eye'', and 39.5\,$\pm$\,3.4~Jy\,km\,s$^{\rm -1}$ in the western 
arm).


\subsubsection{Additional emission lines} \label{subsubsec:other_lines}

Other relevant emission lines in the 3.6~GHz wide band are C$_{\rm 2}$H $N$\,=\,1$-$0, H$^{\rm 13}$CO$^{\rm +}$ $J$\,=\,1$-$0, SiO $J$\,=\,2$-$1. 
C$_{\rm 2}$H has been detected with an integrated flux of 350\,$\pm$\,103~mJy\,km\,s$^{\rm -1}$ (see Fig.\,\ref{fig:3mm_spectrum}). Due to the 
non-detection of H$^{\rm 13}$CO$^{\rm +}$\,1$-$0 and SiO\,2$-$1, only upper limits of $\sim$2.4~mJy\,km\,s$^{\rm -1}$ and 
$\sim$2.5~mJy\,km\,s$^{\rm -1}$, respectively, are available for the fluxes of these line transitions. The limits in the integrated fluxes have been 
obtained from an area corresponding to the largest extension of the HCO$^{\rm +}$ emission, and within a velocity range of $\pm$85~km\,s$^{\rm -1}$ of 
the respective line centers.

\subsection{Interferometric line ratios} \label{subsec:interfero_line_ratios}

The determination of molecular line ratios is a very powerful method to explore the physical and chemical processes in the interstellar medium. However, 
one has to be aware of possible implications when using interferometric data. Indeed, an interferometer is sensitive to a limited range of spatial 
scales. Extended structures might be filtered out if the shortest array baselines are still too long. When comparing two lines, one must therefore ensure 
that the corresponding data are sampling similar spatial scales. Table\,\ref{tab:interfero_info} gives an overview of the properties for each 
observational setup. $\Theta$$_{\rm MRS}$, the maximum recoverable scale, is defined as $\Theta$$_{\rm MRS}$ $\approx$\,0.6\,$\times$\,$\lambda$ / 
B$_{\rm min}$ (e.g., the ALMA Cycle 5 Technical Handbook), where B$_{\rm min}$ is the shortest projected baseline length. For our observations 
$\Theta$$_{\rm MRS}$ ranges between 4.1\arcsec\ for CO\,3$-$2 and 22.0\arcsec\ for HCN\,1$-$0, i.e., significantly larger than the dip of emission in 
CO\,2$-$1 in ``the Eye''. Additionally, we smooth each data set to the same beam size (1.0\arcsec\,$\times$\,1.0\arcsec) for the line ratio 
determination, and we extract the integrated intensities from within the same region.\\
\indent
The $\mathcal{R}_{\rm 32/21}$\,=\,CO\,3$-$2/CO\,2$-$1 brightness temperature ratio is large, $\sim$2.5, at position E, inside the ``Eye of the Medusa'' 
(see Table\,\ref{tab:line_ratios}). The ratios were determined inside a circular 1\arcsec\ polygon. This is slightly larger than the area inside 
`` the Eye'' where CO\,2$-$1 is deficient. Thus, it is very likely that the inner walls of the ``Eye of the Medusa'' contribute to the measured CO\,2$-$1 
flux inside the polygon. Hence, the CO\,3$-$2/CO\,2$-$1 ratio we determined is a lower limit to $\mathcal{R}_{\rm 32/21}$ inside ``the Eye''. To measure 
$\mathcal{R}_{\rm 32/21}$ more precisely, higher-resolution observations are needed.


\section{Discussion} \label{sec:discussion}

\begin{table*}[!ht]
\begin{minipage}[!h]{\textwidth}
\centering
\renewcommand{\footnoterule}{}
\caption{\small
 Brightness temperatures and line widths in the ``Eye of the Medusa'' and the western arm.}
\label{tab:line_ratios}
\tabcolsep0.1cm
\begin{tabular}{lcccccc}
\hline
\noalign{\smallskip}
\hline
\noalign{\smallskip}
Line                 & \multicolumn{3}{c}{``Eye of the Medusa''} & \multicolumn{3}{c}{Western arm}\\
\noalign{\smallskip}
                     & $\int T_{\rm B}\,dv\,$\footnote{The brightness temperatures T$_{\rm B}$ were determined from resolution-matched (1\arcsec) data 
cubes inside the same polygons of 1\arcsec\ size.}\saveFN\Tb\footnote{$\int T_{\rm B}\,dv$ is the brightness temperature integrated over the 1\arcsec\ polygon.}\saveFN\Tbint 
                     & peak $T$$_{\rm B}$\useFN\Tb\footnote{peak T$_{\rm B}$ is the peak brightness temperature of the spectrum integrated over the 1\arcsec\ polygon.}\saveFN\Tbpeak 
                     & FWHM\footnote{Obtained from a single Gaussian fit to the line.}\saveFN\FWHM 
                     & $\int T_{\rm B}\,dv\,$\useFN\Tb \useFN\Tbint
                     & peak $T$$_{\rm B}$\useFN\Tb\useFN\Tbpeak 
                     & FWHM\useFN\FWHM \\
\noalign{\smallskip}
                     & [K\,km/s\,pc$^{\rm 2}$] & [K]  & [km/s] & [K\,km/s\,pc$^{\rm 2}$] & [K] & [km/s]\\
\noalign{\smallskip}
\hline
\noalign{\smallskip}
CO\,3$-$2            & 892.3\,$\pm$\,6.3 & 13.82\,$\pm$\,0.42 & 74.7\,$\pm$\,3.1 & 504.4\,$\pm$\,3.3 & 7.48\,$\pm$\,0.27 & 66.2\,$\pm$\,2.7 \\
CO\,2$-$1            & 356.8\,$\pm$\,0.6 & 3.74\,$\pm$\,0.08  & 88.8\,$\pm$\,2.0 & 208.4\,$\pm$\,0.4 & 2.01\,$\pm$\,0.06 & 91.7\,$\pm$\,2.9 \\
HCO$^{\rm +}$\,1$-$0 & 45.3\,$\pm$\,0.5  & 0.80\,$\pm$\,0.03  & 63.7\,$\pm$\,3.3 & 35.9\,$\pm$\,0.3  & 0.62\,$\pm$\,0.03 & 55.7\,$\pm$\,2.9 \\
HCN\,1$-$0           & 36.4\,$\pm$\,0.2  & 0.53\,$\pm$\,0.03  & 59.6\,$\pm$\,4.2 & 23.7\,$\pm$\,0.3  & 0.51\,$\pm$\,0.04 & 63.6\,$\pm$\,8.6 \\
\noalign{\smallskip}
\hline
\end{tabular}
\end{minipage}
\end{table*}

\begin{figure*}[!ht]
  \centering
    \includegraphics[width=\textwidth]{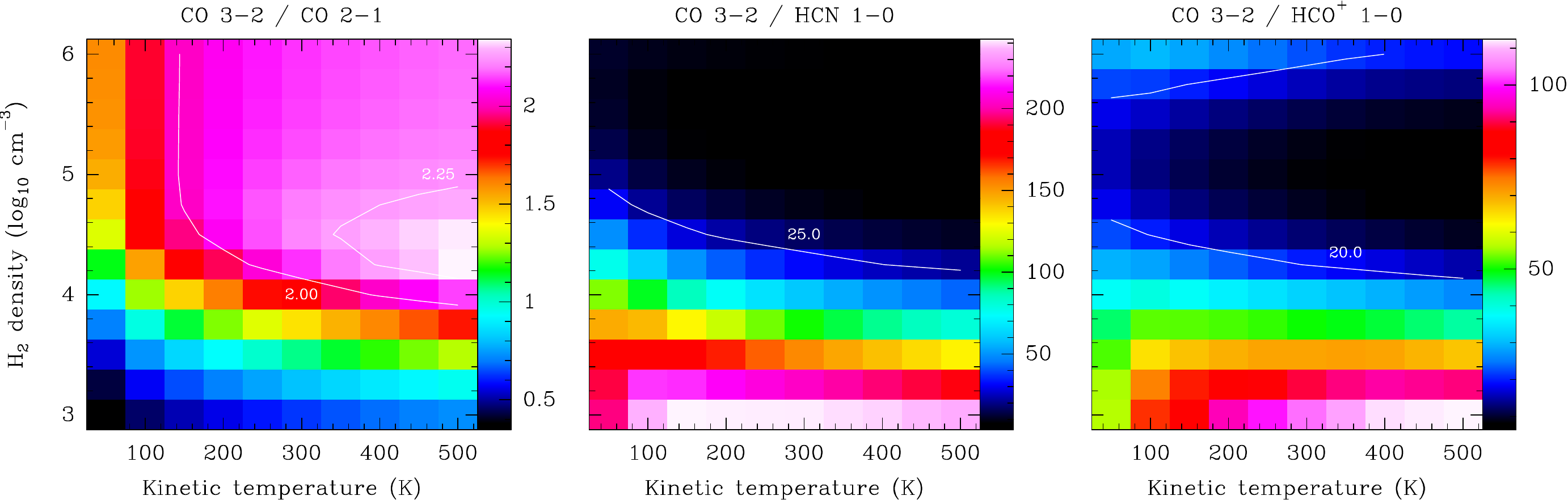}
  \caption{\footnotesize Results of the RADEX analysis of the CO\,3$-$2/CO\,2$-$1, CO\,3$-$2/HCN\,1$-$0, and CO\,3$-$2/HCO$^{\rm +}$\,1$-$0 brightness 
   temperature ratios. The white curves and associated numbers indicate T$_{\rm B}$ ratios toward the ``Eye of the Medusa'' of 2 and 2.25 for 
   CO\,3$-$2/CO\,2$-$1, 25 for CO\,3$-$2/HCN\,1$-$0, and 20 for CO\,3$-$2/HCO$^{\rm +}$\,1$-$0, respectively. The range in H$_{\rm 2}$ densities and 
   kinetic temperatures are the same for each of the three plots. The CO column density (N(CO)/$\Delta$v) is 10$^{\rm 16}$~cm$^{\rm -2}$.} The 
   color scheme indicates possible brightness temperature ratios for each combination of tracers.
  \label{fig:radex}
\end{figure*}

\subsection{Dense gas distribution} \label{subsec:dense_gas_distribution}

A strong correlation between SFR and integrated molecular line luminosity from dense gas is seen in most galaxies \citep[e.g.,][]{gao04b,gao04a}. 
For the Medusa this would imply that due to its high SFR \citep[FIR: 6-7~M$_{\sun}$\,yr$^{\rm -1}$ -- ][H$\alpha$: 
$\sim$10~M$_{\sun}$\,yr$^{\rm -1}$ -- Gallagher in prep.]{aalto00} the HCN luminosity should be high -- that is not the case: On global galaxy scales, 
single-dish observations revealed a low HCN-to-CO\,1$-$0 luminosity ratio similar to values found for GMCs in the galactic disks of \object{M~31} and the 
\object{Milky Way} \citep{bro05,costa11,mat15}, indicating that the fraction of dense gas should be significantly lower than in ULIRGs despite the 
similarly extreme SFE. However, our high-angular resolution observations paint a different picture: the dense gas is not distributed throughout the 
starburst region in the central 2~kpc of the Medusa \citep[e.g.,][]{arm90,wyn93,pre94,aalto00}. Instead, it is located in a compact region inside the 
``Eye of the Medusa'' where the CO\,2$-$1 emission shows a distinct minimum (Fig.\,\ref{fig:hcn1-0+hco+1-0+co3-2}), thus explaining the relative 
faintness on global scales. A comparison between the CO\,2$-$1 emission and the young superstar clusters (SSCs) in the Medusa has shown that many young 
massive clusters are located away from the bulk of the CO \citep{wei04,han06,koenig14} - the same effect is visible for the dense gas, only in a more 
extreme fashion. So far, only one cluster has been identified in the vicinity of the ``Eye'' \citep{han06}. However, the relative astrometric uncertainty 
between the optical and mm data preclude a definite conclusion.\\
\indent
An in-detail kinematic comparison between the HCN, HCO$^{\rm +}$\,1$-$0 and CO\,3$-$2, and the CO\,2$-$1 emission in the ``Eye of the Medusa'' shows 
strong evidence for their different origins. A CO\,2$-$1 spectrum obtained from the center of the ``Eye'' shows a FWHM line width of 
$\sim$90~km\,s$^{\rm -1}$. The HCN and HCO$^{\rm +}$\,1$-$0 spectra obtained from within the same area instead show a FWHM line width of 
$\sim$60~km\,s$^{\rm -1}$ (see Table\,\ref{tab:line_ratios}). The centroid velocities of the lines, however, do not differ significantly (i.e., by 
$\lesssim$10~km\,s$^{\rm -1}$). This may suggest that CO\,2$-$1 is tracing a more extended, lower density component. The CO\,2$-$1 intensity at high 
velocities is roughly the same in the northern and southern part of the ``Eye'', while at lower velocities it is strongly dominated by emission in the 
North. The dense gas emission does not show such a behaviour -- their centroids do not move at all. Thus, the evidence presented here points toward 
different origins of the dense gas and the CO\,2$-$1 emission.

\subsection{Origin and physical properties of the dense gas emission in the ``Eye''} \label{subsec:dense_gas_origin}

\subsubsection{Excitation and line ratios} \label{subsubsec:excitation}

Previous studies, using single-dish as well as interferometric observations, mark the $\mathcal{R}_{\rm 32/21}$ we observed for ``the Eye'' 
($\mathcal{R}_{\rm 32/21}$\,$\sim$2.5, see Sect.\,\ref{subsec:interfero_line_ratios}) as an extreme case: observations of a large number of 
luminous infrared galaxies show that values of $\mathcal{R}_{\rm 32/21}$ greater than unity are rare. So far the highest $\mathcal{R}_{\rm 32/21}$ found 
are at $\sim$1.9 \citep[e.g.,][]{gre09,pap12}. Interferometric observations show that even more powerful ULIRGs, like e.g., \object{Arp~220} and 
\object{NGC~6240}, show $\mathcal{R}_{\rm 32/21}$\,$\lesssim$1 \citep[e.g.,][]{gre09,sli17}. Thus, \textit{ratios measured in the ``Eye of the Medusa'' 
indicate truly exceptional properties of the dense gas}.\\
\indent
$\mathcal{R}_{\rm 32/21}$ together with brightness temperature ratios of CO\,3$-$2 with HCN\,1$-$0 and HCO$^{\rm +}$\,1$-$0 
(Table\,\ref{tab:line_ratios}) can help us to infer the gas properties in this extraordinary region. We perform a line excitation analysis using the 
radiative transfer code RADEX \citep{vdt07} by exploring a grid of parameters in kinetic temperature (50-500~K) and H$_{\rm 2}$ density (10$^{\rm 3}$ to 
10$^{\rm 6}$~cm$^{\rm -3}$) for a CO column density (N(CO)/$\Delta$v) between 10$^{\rm 16}$ and 10$^{\rm 18}$~cm$^{\rm -2}$. Fig.\,\ref{fig:radex} shows 
the results of the radiative transfer modeling: high values ($>$2.2) of $\mathcal{R}_{\rm 32/21}$ can only be achieved for warm gas (T\,$>$300~K), 
H$_{\rm 2}$ densities of a few 10$^{\rm 4}$~cm$^{\rm -3}$ and CO column densities of 10$^{\rm 16}$~cm$^{\rm -2}$. Assuming that HCN, HCO$^{\rm +}$\,1$-$0 
and CO\,3$-$2 trace the same gas component, i.e., the same physical conditions in the gas, we tune the relative abundance ratios for these three 
molecules to match the line ratios for a given density. Adopting also similar beam-filling factors for each tracer, we estimate relative abundances for 
CO/HCN of $\sim$4000 and for CO/HCO$^{\rm +}$ of $\sim$9000. However, since we only have two CO transitions (2$-$1 and 3$-$2) and one transition (1$-$0) 
each of HCN and HCO$^{\rm +}$, the range of possible H$_{\rm 2}$ gas densities is not very well constraint. To change this, observations of additional 
transitions of HCN and/or HCO$^{\rm +}$ are necessary.

\subsubsection{The nature of the dense gas clouds} \label{subsubsec:structure}

As a first step and the simplest scenario, we assume that the dense gas emission originates from one large virialised cloud with a radius of 
$\sim$1\arcsec\ with n(H$_{\rm 2}$)\,$\geq$\,5\,$\times$\,10$^{\rm 3}$~cm$^{\rm -3}$ and T$_{\rm k}$\,$\sim$300~K. This results in an expected linewidth 
(FWHM) of the emission line of about 540~km\,s$^{\rm -1}$. The observed linewidths for all four lines, however, are less than $\sim$100~km\,s$^{\rm -1}$ 
inside ``the Eye'' (see Table\,\ref{tab:line_ratios}). Furthermore, such a structure would have a gas mass of about 10$^{\rm 10}$~M$_{\sun}$ which is 
larger than the gas mass of most galaxies. Thus, the single-cloud, high-filling factor scenario is unrealistic. Alternatively, it's possible that the 
emission is coming from a larger number of small, self-gravitating dense gas clumps. Assuming a typical clump size of 5~pc instead, the scale of dense 
molecular clouds \citep[see e.g.,][]{liz87,ngu11}, results in linewidths of about 10~km\,s$^{\rm -1}$. This scenario is a possibility if the filling 
factor in the gas is high. As a result of this exercise, it seems that the dense gas in the ``Eye of the Medusa'' is most likely fragmented into a large 
number of small filaments or clouds.

\subsection{What is inside the ``Eye of the Medusa''?} \label{subsec:what_is_the_eye}

\subsubsection{Spectral index} \label{subsubsec:spectral_index}

Continuum observations in the radio and mm regimes can provide hints towards the excitation mechanism for the dense gas emission. For the Medusa, 
high-resolution continuum data at 150~MHz (LOFAR, K\"onig et al. in prep.), 1.4, 5, 8.4, 15 and 22~GHz 
\citep[eMERLIN, VLA,][König et al. in prep.]{beck14}, and at 86, 230 and 345~GHz \citep[this work,][]{koenig14} show spectral indices indicative of a 
mixture of non-thermal synchrotron and thermal free-free emission. These emission processes are usually associated with supernovae (SNe) and AGN 
activity (synchrotron) and young star formation \citep[free-free, e.g.,][and references therein]{beck14,scho17}.\\
\indent
In an earlier paper, we have shown that it is possible that the energetic output from a large number of supernova explosions could have led to the 
observed shell-like morphology of the CO\,2$-$1 emission in ``the Eye of the Medusa'' \citep{koenig14}. FIR and X-ray observations hint towards the 
presence of low-level AGN activity at the dynamic center of the Medusa \citep[e.g.,][]{kaa08,ber09,leh10}. The spatially resolved studies 
\citep{kaa08,leh10} found the activity to be associated with the main 1.4~GHz radio continuum emission peak north of ``the Eye'' \citep{bes05}. So far 
no indications for an AGN inside the ``Eye of the Medusa'' have been found. Thus, supernovae are the most likely contributor of the synchrotron emission 
in ``the Eye''.\\
\indent
The free-free component contributing to the continuum emission could be due to the presence of young star clusters inside ``the Eye'', which could be 
heavily obscured \citep{han06}. However, the faint mm continuum emission towards higher frequencies for ``the Eye'' (see Sect.\,\ref{subsec:cont}) 
indicates that there is very little cold dust emission associated with the dense gas in this region. The high gas temperature (see 
Sect.\,\ref{subsubsec:excitation}) most likely leads to the dust being heated and thus its emission peak is shifted to shorter thermal infrared 
wavelengths. Observations at these wavelengths, that would provide a test of this hypothesis, are currently not available for the Medusa.\\
\indent
The mixed spectral index indicates that the age of the star formation in ``the Eye'' would be at least a few 10$^{\rm 6}$~yrs, since at this stage in 
massive stellar evolution the first supernovae occur.

\subsubsection{HCN vs. HCO$^{\rm +}$} \label{subsubsec:line_ratios}

Single-dish observations of the total molecular gas and the dense molecular gas on large scales have shown differences in the global properties in a 
range of activity types in galaxies \citep[e.g.,][]{sol97,gra08,kri08,pri15}. Several studies have found intrinsic differences between the way dense 
molecular gas content is traced by e.g., HCN and HCO$^{\rm +}$, and the distribution of low-$J$ CO transitions as tracers of the total gas properties: 
$\mathcal{R}_{\rm HCN/CO}$, the HCN-to-CO ratio, is small in quiescent and starburst galaxies 
\citep[$<$0.3,][and references therein]{gao04a,kri07,mat15} compared to galaxies with AGN activity 
\citep[up to 1 or more,][and references therein]{use04,kri07,mat15}. On small scales $\mathcal{R}_{\rm HCN/CO}$ can be significantly different from 
values found on galaxy scales \citep[e.g., in \object{NGC~1068}, \object{NGC~6951}, and \object{M~51}][]{mat15,kri07,koh96,kri11}. Our 
high-resolution ($\sim$1\arcsec) HCN\,1$-$0-to-CO\,2$-$1 line ratio is $\sim$0.20 in the ``Eye of the Medusa''. An interferometric CO\,2$-$1-to-CO\,1$-$0 
line ratio map yields values of $\sim$0.9 in ``the Eye'' (K\"onig et al., in prep.). Taking the CO ratio into account, the HCN\,1$-$0-to-CO\,2$-$1 ratio 
translates to HCN\,1$-$0-to-CO\,1$-$0 of $\sim$0.1. This places ``the Eye'' in a regime close to what has been found LIRGs and ULIRGs 
\citep[e.g.,][]{gao07,use15}.\\
\indent
It has also been shown that $\mathcal{R}_{\rm HCN/HCO^{+}}$, the HCN-to-HCO$^{\rm +}$ line ratio, can be significantly higher close to AGN (nuclear) 
compared to regions dominated by starbursts, and regions of more quiescent star formation \citep[e.g.,][]{gra06,gra08,kri08,koh08,dav12}, though it can 
also be high in young (pre-synchrotron) star forming regions. $\mathcal{R}_{\rm HCN/HCO^{+}}$ inside ``the Eye'' is approximately 0.9, which would 
place the ``Eye of the Medusa'' in the regime of starburst galaxies.

\subsubsection{CN as a tracer of radiative feedback} \label{subsubsec:cn}

The reasons for the enhancement of HCN and/or HCO$^{\rm +}$ relative to CO and/or each other are diverse: strong UV/X-ray radiation fields 
\citep[e.g.,][]{use04,mei06,mei07,kri07,kri11,mat15}, energetic particles from AGN and/or jets, cosmic rays 
\citep[e.g.,][and references therein]{mei06,bay10}, young star formation \citep[e.g.,][and references therein]{pir99,boo01,lah07}, IR-pumping of HCN 
\citep[e.g.,][]{car81,aalto95,mat15}, high-temperature gas chemistry. A molecule that can help differentiate between these scenarios is CN. This 
molecule can be formed from HCN as a result of intense UV radiation in photodissociation regions \citep[PDRs, e.g.,][]{aalto02,baan08,han15}. 
\citet{costa11} found that in the Medusa, globally CN\,1$-$0 is found at higher intensity than HCN\,1$-$0. If we assume that the CN and HCN emission 
originates from the same spatial region inside the ``Eye of the Medusa'', the CN-to-HCN\,1$-$0 brightness temperature ratio is $\sim$2.1. This might 
imply that a large part of the HCN was photodissociated into CN due to the intense UV radiation from nearby young massive stars.

\subsubsection{Scenarios} \label{subsubsec:scenarios}

What we know so far about the properties of the central ISM in the Medusa is that the gas inside ``the Eye'' is hot and dense, that there is very little 
cold dust, and that a mixture of synchrotron and free-free emission processes take places in this region. What exactly is going on in this region is not 
clearly determined, yet. So far, the evidence seems to point towards post-burst radiative feedback from young massive stars as the energetic source of 
the observed phenomena. In this scenario, HCN and HCO$^{\rm +}$\,1$-$0, and CO\,3$-$2 are enhanced in the dense, hot gas that is associated with deeply 
embedded, fragmenting star-formation cores.\\
\indent
Alternatively however, the dense gas properties could also be the immediate result of shocks, in which the gas is compressed and heated. CO studies have 
shown that shock-excited molecular gas can reach temperatures and densities similar, or above, to the values we find for the ``Eye of the Medusa'' 
\citep[e.g., \object{NGC~1068}, \object{NGC~6240}, \object{NGC~7130},][]{hai12,mei13,poz17}.\\
\indent
To be able to distinguish between the two scenarios and better determine the small-scale structure of the dense gas in ``the Eye'', further observations are necessary. Higher-resolution data sets of dense gas tracers, like the ones discussed in this paper, will help determine the degree of fragmentation 
in the ``Eye'' ISM, and to track the locations of shocks and/or clouds in which the massive stars are embedded. Observations of tracers such as SiO, 
CH$_{\rm 3}$OH at millimeter wavelengths, and NIR H$_{\rm 2}$ will directly reveal the presence of shocks, whereas CN observations will reveal UV-photon 
dominated regions close to stars. Also, shocks do not heat dust as effectively as they heat gas \citep[e.g.,][]{mei13}. Thus a good test will be to 
determine the conditions in the warm dust at infrared wavelengths.

\subsection{The importance of interaction-induced gas flows} \label{subsec:other_galaxies}

Studies have shown that the global CO-to-HCN ratio does not necessarily reflect on the small-scale properties of the dense gas, i.e., a global value is 
not representative of the small-scale processes leading to the observed ratios \citep[e.g., in \object{NGC~1068}, \object{NGC~6951}, and 
\object{M~51}][]{mat15,kri07,koh96,kri11}. For the Medusa, its relatively high global CO-to-HCN ratio implies HCN, and thus the dense gas, to be faint. 
This is puzzling however, especially when taking into account the Medusa's high star formation efficiency. Following \citet{ken98} and 
\citet{gao04a,gao04b}, one would expect the known star-forming complexes to be associated with dense molecular gas. Our high-resolution observations 
show, however, that the dense gas is located in a very compact configuration inside the ``Eye of the Medusa'', away from most of the optically visible 
on-going star formation and the bulk of the molecular gas. However, the 3~mm continuum associated with the dense gas inside ``the Eye'' yields a SFR of 
about 2.8~M$_{\sun}$\,yr$^{\rm -1}$ \citep[following][]{mur11}. Thus, obscured star formation seems to be on-going in this region. An important factor 
contributing to this puzzle could be the presence of inflowing gas in this minor merger.\\
\indent
It has been shown that the inflow of molecular gas in galaxy mergers or interactions can lead to an increase in and/or the triggering of star formation 
\citep[e.g.,][and references therein]{mei02,whi07,tur15}. The Medusa is suspected to harbor a molecular gas inflow via its kpc-scale dust lane to the 
nucleus \citep{aalto10,koenig14}, similar to what has been found in e.g., NGC~5253 \citep{tur15}. For NGC~5253 it has been suggested that it has 
experienced an encounter with another galaxy in its history. Observations of the dense molecular gas, traced by CO\,3$-$2, in this dwarf galaxy led the 
authors to conclude that the increased star formation activity they find in a gas cloud previously only weakly detected in CO\,2$-$1 is the result of 
gas inflow. Molecular gas is being transported along the minor axis dust lane, east of the nucleus, towards the galaxy center (the ``streamer''). 
Associated with the CO\,3$-$2-detection is an ``extremely'' dusty gas cloud that contains two very young stellar clusters of a few times 
10$^{\rm 5}$~M$_{\sun}$ each \citep{cal15}. The star formation inside these clusters was likely triggered as a result of the interaction of the 
inflowing gas with the prevailing ISM \citep{tur15}.\\
\indent
This scenario is a strong possibility for NGC~4194 as well. However, the affected region in the ``Eye of the Medusa'' ($r$\,$\sim$185~pc) is much larger 
(about 100 times the area) than what has been found in NGC~5253 ($\sim$31~pc\,$\times$\,11~pc) -- enough space to host several star clusters. Also, in 
contrast to NGC~5253, very little cold dust emission is associated with the dense gas in ``the Eye'' (see Sect.\,\ref{subsubsec:spectral_index}). 
Furthermore, only a small amount of the dense gas emission in the Medusa is found in the dust lane associated with the gas inflow (see 
Figs.\,\ref{fig:co2-1+3mmcont}, \ref{fig:hcn1-0+hco+1-0+co3-2}). Whether these differences could be due to the dissimilar size scales in NGC~5253 and the 
Medusa, or a difference in the time scales of the processes induced by the gas inflow, is unclear.


\section{Conclusions} \label{sec:conclusions}

In this paper, we studied the distribution and properties of the dense gas in the Medusa merger through observations of HCN and HCO$^{\rm +}$\,1$-$0 and 
CO\,3$-$2. Surprisingly, all these tracers are located inside the ``Eye of the Medusa'', where CO\,2$-$1 emission shows a distinct minimum. The gas 
inside ``the Eye'' is hot ($>$300~K) and dense ($\gtrsim$10$^{\rm 4}$~cm$^{\rm -3}$). We propose two possible scenarios to explain what could cause the 
extreme properties of the gas inside the ``Eye of the Medusa'': 1) they are caused by shocks as the gas flow, that transports molecular gas via the dust 
lane to the center of the Medusa, collides with the central ISM inside ``the Eye'', or 2) they are an effect of supernova-induced shocks and/or radiative 
feedback from embedded massive star clusters, where the star formation is fed by the gas flow. Additional higher resolution observations 
($\sim$0.5\arcsec) of e.g., HCN and HCO$^{\rm +}$\,3$-$2, and CN\,2$-$1, and H$_{\rm 2}$ NIR observations will help us disentangle the energetic 
processes in the ``Eye of the Medusa'' to determine what caused the extreme properties in the dense gas.


\begin{acknowledgements}

We thank the referee for useful comments. A.A. acknowledges the support of the Swedish Research Council (Vetenskapsr\r{a}det) and the Swedish National 
Space Board (SNSB). The Submillimeter Array is a joint project between the Smithsonian Astrophysical Observatory and the Academia Sinica Institute of 
Astronomy and Astrophysics and is funded by the Smithsonian Institution and the Academia Sinica. This paper is based in part on data obtained with the 
International LOFAR Telescope (ILT) under project code LC7\_006. LOFAR \citep{vanH13} is the Low Frequency Array designed and constructed by ASTRON. It 
has observing, data processing, and data storage facilities in several countries, that are owned by various parties (each with their own funding 
sources), and that are collectively operated by the ILT foundation under a joint scientific policy. The ILT resources have benefitted from the following 
recent major funding sources: CNRS-INSU, Observatoire de Paris and Universit\'{e} d'Orl\'{e}ans, France; BMBF, MIWF-NRW, MPG, Germany; Science Foundation 
Ireland (SFI), Department of Business, Enterprise and Innovation (DBEI), Ireland; NWO, The Netherlands; The Science and Technology Facilities Council, 
UK. MERLIN/eMERLIN is a National Facility operated by the University of Manchester at Jodrell Bank Observatory on behalf of STFC. This research has made 
use of the NASA/IPAC Extragalactic Database (NED) which is operated by the Jet Propulsion Laboratory, California Institute of Technology, under contract 
with the National Aeronautics and Space Administration.
\end{acknowledgements}


\bibliographystyle{aa}
\bibliography{ngc4194_hcnhco+}

\begin{thebibliography}{74}
\expandafter\ifx\csname natexlab\endcsname\relax\def\natexlab#1{#1}\fi

\bibitem[{{Aalto} {et~al.}(2010){Aalto}, {Beswick}, \& {J{\"u}tte}}]{aalto10}
{Aalto}, S., {Beswick}, R., \& {J{\"u}tte}, E. 2010, \aap, 522, A59

\bibitem[{{Aalto} {et~al.}(1995){Aalto}, {Booth}, {Black}, \&
  {Johansson}}]{aalto95}
{Aalto}, S., {Booth}, R.~S., {Black}, J.~H., \& {Johansson}, L.~E.~B. 1995,
  \aap, 300, 369

\bibitem[{{Aalto} \& {H{\"u}ttemeister}(2000)}]{aalto00}
{Aalto}, S. \& {H{\"u}ttemeister}, S. 2000, \aap, 362, 42

\bibitem[{{Aalto} {et~al.}(2001){Aalto}, {H{\"u}ttemeister}, \&
  {Polatidis}}]{aalto01}
{Aalto}, S., {H{\"u}ttemeister}, S., \& {Polatidis}, A.~G. 2001, \aap, 372, L29

\bibitem[{{Aalto} {et~al.}(2002){Aalto}, {Polatidis}, {H{\"u}ttemeister}, \&
  {Curran}}]{aalto02}
{Aalto}, S., {Polatidis}, A.~G., {H{\"u}ttemeister}, S., \& {Curran}, S.~J.
  2002, \aap, 381, 783

\bibitem[{{Armus} {et~al.}(1990){Armus}, {Heckman}, \& {Miley}}]{arm90}
{Armus}, L., {Heckman}, T.~M., \& {Miley}, G.~K. 1990, \apj, 364, 471

\bibitem[{{Baan} {et~al.}(2008){Baan}, {Henkel}, {Loenen}, {Baudry}, \&
  {Wiklind}}]{baan08}
{Baan}, W.~A., {Henkel}, C., {Loenen}, A.~F., {Baudry}, A., \& {Wiklind}, T.
  2008, \aap, 477, 747

\bibitem[{{Bayet} {et~al.}(2010){Bayet}, {Hartquist}, {Viti}, {Williams}, \&
  {Bell}}]{bay10}
{Bayet}, E., {Hartquist}, T.~W., {Viti}, S., {Williams}, D.~A., \& {Bell},
  T.~A. 2010, \aap, 521, A16

\bibitem[{{Beck} {et~al.}(2014){Beck}, {Lacy}, {Turner}, {Greathouse}, \&
  {Neff}}]{beck14}
{Beck}, S.~C., {Lacy}, J., {Turner}, J., {Greathouse}, T., \& {Neff}, S. 2014,
  \apj, 787, 85

\bibitem[{{Bernard-Salas} {et~al.}(2009){Bernard-Salas}, {Spoon},
  {Charmandaris}, {Lebouteiller}, {Farrah}, {Devost}, {Brandl}, {Wu}, {Armus},
  {Hao}, {Sloan}, {Weedman}, \& {Houck}}]{ber09}
{Bernard-Salas}, J., {Spoon}, H.~W.~W., {Charmandaris}, V., {et~al.} 2009,
  \apjs, 184, 230

\bibitem[{{Beswick} {et~al.}(2005){Beswick}, {Aalto}, {Pedlar}, \&
  {H{\"u}ttemeister}}]{bes05}
{Beswick}, R.~J., {Aalto}, S., {Pedlar}, A., \& {H{\"u}ttemeister}, S. 2005,
  \aap, 444, 791

\bibitem[{{Bonatto} {et~al.}(1999){Bonatto}, {Bica}, {Pastoriza}, \&
  {Alloin}}]{bon99}
{Bonatto}, C., {Bica}, E., {Pastoriza}, M.~G., \& {Alloin}, D. 1999, \aap, 343,
  100

\bibitem[{{Boonman} {et~al.}(2001){Boonman}, {Stark}, {van der Tak}, {van
  Dishoeck}, {van der Wal}, {Sch{\"a}fer}, {de Lange}, \& {Laauwen}}]{boo01}
{Boonman}, A.~M.~S., {Stark}, R., {van der Tak}, F.~F.~S., {et~al.} 2001,
  \apjl, 553, L63

\bibitem[{{Brouillet} {et~al.}(2005){Brouillet}, {Muller}, {Herpin}, {Braine},
  \& {Jacq}}]{bro05}
{Brouillet}, N., {Muller}, S., {Herpin}, F., {Braine}, J., \& {Jacq}, T. 2005,
  \aap, 429, 153

\bibitem[{{Calzetti} {et~al.}(2015){Calzetti}, {Johnson}, {Adamo}, {Gallagher},
  {Andrews}, {Smith}, {Clayton}, {Lee}, {Sabbi}, {Ubeda}, {Kim}, {Ryon},
  {Thilker}, {Bright}, {Zackrisson}, {Kennicutt}, {de Mink}, {Whitmore},
  {Aloisi}, {Chandar}, {Cignoni}, {Cook}, {Dale}, {Elmegreen}, {Elmegreen},
  {Evans}, {Fumagalli}, {Gouliermis}, {Grasha}, {Grebel}, {Krumholz},
  {Walterbos}, {Wofford}, {Brown}, {Christian}, {Dobbs}, {Herrero}, {Kahre},
  {Messa}, {Nair}, {Nota}, {{\"O}stlin}, {Pellerin}, {Sacchi}, {Schaerer}, \&
  {Tosi}}]{cal15}
{Calzetti}, D., {Johnson}, K.~E., {Adamo}, A., {et~al.} 2015, \apj, 811, 75

\bibitem[{{Carroll} \& {Goldsmith}(1981)}]{car81}
{Carroll}, T.~J. \& {Goldsmith}, P.~F. 1981, \apj, 245, 891

\bibitem[{{Costagliola} {et~al.}(2011){Costagliola}, {Aalto}, {Rodriguez},
  {Muller}, {Spoon}, {Mart{\'{\i}}n}, {Per{\'e}z-Torres}, {Alberdi},
  {Lindberg}, {Batejat}, {J{\"u}tte}, {van der Werf}, \& {Lahuis}}]{costa11}
{Costagliola}, F., {Aalto}, S., {Rodriguez}, M.~I., {et~al.} 2011, \aap, 528,
  A30

\bibitem[{{Curran} {et~al.}(2000){Curran}, {Aalto}, \& {Booth}}]{cur00}
{Curran}, S.~J., {Aalto}, S., \& {Booth}, R.~S. 2000, \aaps, 141, 193

\bibitem[{{Davies} {et~al.}(2012){Davies}, {Mark}, \& {Sternberg}}]{dav12}
{Davies}, R., {Mark}, D., \& {Sternberg}, A. 2012, \aap, 537, A133

\bibitem[{{Gao} {et~al.}(2007){Gao}, {Carilli}, {Solomon}, \& {Vanden
  Bout}}]{gao07}
{Gao}, Y., {Carilli}, C.~L., {Solomon}, P.~M., \& {Vanden Bout}, P.~A. 2007,
  \apjl, 660, L93

\bibitem[{{Gao} \& {Solomon}(2004{\natexlab{a}})}]{gao04b}
{Gao}, Y. \& {Solomon}, P.~M. 2004{\natexlab{a}}, \apjs, 152, 63

\bibitem[{{Gao} \& {Solomon}(2004{\natexlab{b}})}]{gao04a}
{Gao}, Y. \& {Solomon}, P.~M. 2004{\natexlab{b}}, \apj, 606, 271

\bibitem[{{Graci{\'a}-Carpio} {et~al.}(2006){Graci{\'a}-Carpio},
  {Garc{\'{\i}}a-Burillo}, {Planesas}, \& {Colina}}]{gra06}
{Graci{\'a}-Carpio}, J., {Garc{\'{\i}}a-Burillo}, S., {Planesas}, P., \&
  {Colina}, L. 2006, \apjl, 640, L135

\bibitem[{{Graci{\'a}-Carpio} {et~al.}(2008){Graci{\'a}-Carpio},
  {Garc{\'{\i}}a-Burillo}, {Planesas}, {Fuente}, \& {Usero}}]{gra08}
{Graci{\'a}-Carpio}, J., {Garc{\'{\i}}a-Burillo}, S., {Planesas}, P., {Fuente},
  A., \& {Usero}, A. 2008, \aap, 479, 703

\bibitem[{{Greve} {et~al.}(2009){Greve}, {Papadopoulos}, {Gao}, \&
  {Radford}}]{gre09}
{Greve}, T.~R., {Papadopoulos}, P.~P., {Gao}, Y., \& {Radford}, S.~J.~E. 2009,
  \apj, 692, 1432

\bibitem[{{Hailey-Dunsheath} {et~al.}(2012){Hailey-Dunsheath}, {Sturm},
  {Fischer}, {Sternberg}, {Graci{\'a}-Carpio}, {Davies},
  {Gonz{\'a}lez-Alfonso}, {Mark}, {Poglitsch}, {Contursi}, {Genzel}, {Lutz},
  {Tacconi}, {Veilleux}, {Verma}, \& {de Jong}}]{hai12}
{Hailey-Dunsheath}, S., {Sturm}, E., {Fischer}, J., {et~al.} 2012, \apj, 755,
  57

\bibitem[{{Han} {et~al.}(2015){Han}, {Zhou}, {Wang}, {Esimbek}, {Zhang}, \&
  {Wang}}]{han15}
{Han}, X.~H., {Zhou}, J.~J., {Wang}, J.~Z., {et~al.} 2015, \aap, 576, A131

\bibitem[{{Hancock} {et~al.}(2006){Hancock}, {Weistrop}, {Nelson}, \&
  {Kaiser}}]{han06}
{Hancock}, M., {Weistrop}, D., {Nelson}, C.~H., \& {Kaiser}, M.~E. 2006, \aj,
  131, 282

\bibitem[{{Hernquist} \& {Mihos}(1995)}]{her95}
{Hernquist}, L. \& {Mihos}, J.~C. 1995, \apj, 448, 41

\bibitem[{{Kaaret} \& {Alonso-Herrero}(2008)}]{kaa08}
{Kaaret}, P. \& {Alonso-Herrero}, A. 2008, \apj, 682, 1020

\bibitem[{{Kaviraj}(2014)}]{kav14}
{Kaviraj}, S. 2014, \mnras, 440, 2944

\bibitem[{{Kaviraj}(2016)}]{kav16}
{Kaviraj}, S. 2016, in IAU Symposium, Vol. 319, Galaxies at High Redshift and
  Their Evolution Over Cosmic Time, ed. S.~{Kaviraj}, 130--136

\bibitem[{{Kaviraj} {et~al.}(2009){Kaviraj}, {Peirani}, {Khochfar}, {Silk}, \&
  {Kay}}]{kav09}
{Kaviraj}, S., {Peirani}, S., {Khochfar}, S., {Silk}, J., \& {Kay}, S. 2009,
  \mnras, 394, 1713

\bibitem[{{Kennicutt}(1998)}]{ken98}
{Kennicutt}, Jr., R.~C. 1998, \apj, 498, 541

\bibitem[{{Kohno} {et~al.}(1996){Kohno}, {Kawabe}, {Tosaki}, \&
  {Okumura}}]{koh96}
{Kohno}, K., {Kawabe}, R., {Tosaki}, T., \& {Okumura}, S.~K. 1996, \apjl, 461,
  L29

\bibitem[{{Kohno} {et~al.}(2008){Kohno}, {Nakanishi}, {Tosaki}, {Muraoka},
  {Miura}, {Ezawa}, \& {Kawabe}}]{koh08}
{Kohno}, K., {Nakanishi}, K., {Tosaki}, T., {et~al.} 2008, \apss, 313, 279

\bibitem[{{K{\"o}nig} {et~al.}(2014){K{\"o}nig}, {Aalto}, {Lindroos}, {Muller},
  {Gallagher}, {Beswick}, {Petitpas}, \& {J{\"u}tte}}]{koenig14}
{K{\"o}nig}, S., {Aalto}, S., {Lindroos}, L., {et~al.} 2014, \aap, 569, A6

\bibitem[{{Kormendy} \& {Cornell}(2004)}]{kor04}
{Kormendy}, J. \& {Cornell}, M.~E. 2004, in Astrophysics and Space Science
  Library, Vol. 319, Penetrating Bars Through Masks of Cosmic Dust, ed. D.~L.
  {Block}, I.~{Puerari}, K.~C. {Freeman}, R.~{Groess}, \& E.~K. {Block}, 261

\bibitem[{{Krips} {et~al.}(2011){Krips}, {Mart{\'{\i}}n}, {Eckart}, {Neri},
  {Garc{\'{\i}}a-Burillo}, {Matsushita}, {Peck}, {Stoklasov{\'a}}, {Petitpas},
  {Usero}, {Combes}, {Schinnerer}, {Humphreys}, \& {Baker}}]{kri11}
{Krips}, M., {Mart{\'{\i}}n}, S., {Eckart}, A., {et~al.} 2011, \apj, 736, 37

\bibitem[{{Krips} {et~al.}(2007){Krips}, {Neri}, {Garc{\'{\i}}a-Burillo},
  {Combes}, {Schinnerer}, {Baker}, {Eckart}, {Boone}, {Hunt}, {Leon}, \&
  {Tacconi}}]{kri07}
{Krips}, M., {Neri}, R., {Garc{\'{\i}}a-Burillo}, S., {et~al.} 2007, \aap, 468,
  L63

\bibitem[{{Krips} {et~al.}(2008){Krips}, {Neri}, {Garc{\'{\i}}a-Burillo},
  {Mart{\'{\i}}n}, {Combes}, {Graci{\'a}-Carpio}, \& {Eckart}}]{kri08}
{Krips}, M., {Neri}, R., {Garc{\'{\i}}a-Burillo}, S., {et~al.} 2008, \apj, 677,
  262

\bibitem[{{Lahuis} {et~al.}(2007){Lahuis}, {Spoon}, {Tielens}, {Doty}, {Armus},
  {Charmandaris}, {Houck}, {St{\"a}uber}, \& {van Dishoeck}}]{lah07}
{Lahuis}, F., {Spoon}, H.~W.~W., {Tielens}, A.~G.~G.~M., {et~al.} 2007, \apj,
  659, 296

\bibitem[{{Lehmer} {et~al.}(2010){Lehmer}, {Alexander}, {Bauer}, {Brandt},
  {Goulding}, {Jenkins}, {Ptak}, \& {Roberts}}]{leh10}
{Lehmer}, B.~D., {Alexander}, D.~M., {Bauer}, F.~E., {et~al.} 2010, \apj, 724,
  559

\bibitem[{{Lizano} \& {Shu}(1987)}]{liz87}
{Lizano}, S. \& {Shu}, F.~H. 1987, in NATO ASIC Proc. 210: Physical Processes
  in Interstellar Clouds, ed. G.~E. {Morfill} \& M.~{Scholer}, 173--193

\bibitem[{{Lotz} {et~al.}(2011){Lotz}, {Jonsson}, {Cox}, {Croton}, {Primack},
  {Somerville}, \& {Stewart}}]{lot11}
{Lotz}, J.~M., {Jonsson}, P., {Cox}, T.~J., {et~al.} 2011, \apj, 742, 103

\bibitem[{{Matsushita} {et~al.}(2015){Matsushita}, {Trung}, {Boone}, {Krips},
  {Lim}, \& {Muller}}]{mat15}
{Matsushita}, S., {Trung}, D.-V., {Boone}, F., {et~al.} 2015, \apj, 799, 26

\bibitem[{{Meier} {et~al.}(2002){Meier}, {Turner}, \& {Beck}}]{mei02}
{Meier}, D.~S., {Turner}, J.~L., \& {Beck}, S.~C. 2002, \aj, 124, 877

\bibitem[{{Meijerink} {et~al.}(2013){Meijerink}, {Kristensen}, {Wei{\ss}}, {van
  der Werf}, {Walter}, {Spaans}, {Loenen}, {Fischer}, {Israel}, {Isaak},
  {Papadopoulos}, {Aalto}, {Armus}, {Charmandaris}, {Dasyra}, {Diaz-Santos},
  {Evans}, {Gao}, {Gonz{\'a}lez-Alfonso}, {G{\"u}sten}, {Henkel}, {Kramer},
  {Lord}, {Mart{\'{\i}}n-Pintado}, {Naylor}, {Sanders}, {Smith}, {Spinoglio},
  {Stacey}, {Veilleux}, \& {Wiedner}}]{mei13}
{Meijerink}, R., {Kristensen}, L.~E., {Wei{\ss}}, A., {et~al.} 2013, \apjl,
  762, L16

\bibitem[{{Meijerink} {et~al.}(2006){Meijerink}, {Spaans}, \& {Israel}}]{mei06}
{Meijerink}, R., {Spaans}, M., \& {Israel}, F.~P. 2006, \apjl, 650, L103

\bibitem[{{Meijerink} {et~al.}(2007){Meijerink}, {Spaans}, \& {Israel}}]{mei07}
{Meijerink}, R., {Spaans}, M., \& {Israel}, F.~P. 2007, \aap, 461, 793

\bibitem[{{Murphy} {et~al.}(2011){Murphy}, {Condon}, {Schinnerer}, {Kennicutt},
  {Calzetti}, {Armus}, {Helou}, {Turner}, {Aniano}, {Beir{\~a}o}, {Bolatto},
  {Brandl}, {Croxall}, {Dale}, {Donovan Meyer}, {Draine}, {Engelbracht},
  {Hunt}, {Hao}, {Koda}, {Roussel}, {Skibba}, \& {Smith}}]{mur11}
{Murphy}, E.~J., {Condon}, J.~J., {Schinnerer}, E., {et~al.} 2011, \apj, 737,
  67

\bibitem[{{Nguyen Luong} {et~al.}(2011){Nguyen Luong}, {Motte}, {Schuller},
  {Schneider}, {Bontemps}, {Schilke}, {Menten}, {Heitsch}, {Wyrowski},
  {Carlhoff}, {Bronfman}, \& {Henning}}]{ngu11}
{Nguyen Luong}, Q., {Motte}, F., {Schuller}, F., {et~al.} 2011, \aap, 529, A41

\bibitem[{{Papadopoulos} {et~al.}(2012){Papadopoulos}, {van der Werf},
  {Xilouris}, {Isaak}, {Gao}, \& {M{\"u}hle}}]{pap12}
{Papadopoulos}, P.~P., {van der Werf}, P.~P., {Xilouris}, E.~M., {et~al.} 2012,
  \mnras, 426, 2601

\bibitem[{{Pellerin} \& {Robert}(2007)}]{pel07}
{Pellerin}, A. \& {Robert}, C. 2007, \mnras, 381, 228

\bibitem[{{Pirogov}(1999)}]{pir99}
{Pirogov}, L. 1999, \aap, 348, 600

\bibitem[{{Pozzi} {et~al.}(2017){Pozzi}, {Vallini}, {Vignali}, {Talia},
  {Gruppioni}, {Mingozzi}, {Massardi}, \& {Andreani}}]{poz17}
{Pozzi}, F., {Vallini}, L., {Vignali}, C., {et~al.} 2017, \mnras, 470, L64

\bibitem[{{Prestwich} {et~al.}(1994){Prestwich}, {Joseph}, \& {Wright}}]{pre94}
{Prestwich}, A.~H., {Joseph}, R.~D., \& {Wright}, G.~S. 1994, \apj, 422, 73

\bibitem[{{Privon} {et~al.}(2015){Privon}, {Herrero-Illana}, {Evans},
  {Iwasawa}, {Perez-Torres}, {Armus}, {D{\'{\i}}az-Santos}, {Murphy},
  {Stierwalt}, {Aalto}, {Mazzarella}, {Barcos-Mu{\~n}oz}, {Borish}, {Inami},
  {Kim}, {Treister}, {Surace}, {Lord}, {Conway}, {Frayer}, \&
  {Alberdi}}]{pri15}
{Privon}, G.~C., {Herrero-Illana}, R., {Evans}, A.~S., {et~al.} 2015, \apj,
  814, 39

\bibitem[{{Sandage}(1990)}]{sandage90}
{Sandage}, A. 1990, \jrasc, 84, 70

\bibitem[{{Sandage}(2005)}]{sandage05}
{Sandage}, A. 2005, \araa, 43, 581

\bibitem[{{Schmidt}(1959)}]{schm59}
{Schmidt}, M. 1959, \apj, 129, 243

\bibitem[{{Schober} {et~al.}(2017){Schober}, {Schleicher}, \&
  {Klessen}}]{scho17}
{Schober}, J., {Schleicher}, D.~R.~G., \& {Klessen}, R.~S. 2017, \mnras, 468,
  946

\bibitem[{{Shlosman}(2013)}]{shl13}
{Shlosman}, I. 2013, {Cosmological Evolution of Galaxies}, ed.
  J.~{Falc{\'o}n-Barroso} \& J.~H. {Knapen}, 555

\bibitem[{{Sliwa} \& {Downes}(2017)}]{sli17}
{Sliwa}, K. \& {Downes}, D. 2017, \aap, 604, A2

\bibitem[{{Solomon} {et~al.}(1992){Solomon}, {Downes}, \& {Radford}}]{sol92}
{Solomon}, P.~M., {Downes}, D., \& {Radford}, S.~J.~E. 1992, \apjl, 387, L55

\bibitem[{{Solomon} {et~al.}(1997){Solomon}, {Downes}, {Radford}, \&
  {Barrett}}]{sol97}
{Solomon}, P.~M., {Downes}, D., {Radford}, S.~J.~E., \& {Barrett}, J.~W. 1997,
  \apj, 478, 144

\bibitem[{{Turner} {et~al.}(2015){Turner}, {Beck}, {Benford}, {Consiglio},
  {Ho}, {Kov{\'a}cs}, {Meier}, \& {Zhao}}]{tur15}
{Turner}, J.~L., {Beck}, S.~C., {Benford}, D.~J., {et~al.} 2015, \nat, 519, 331

\bibitem[{{Usero} {et~al.}(2004){Usero}, {Garc{\'{\i}}a-Burillo}, {Fuente},
  {Mart{\'{\i}}n-Pintado}, \& {Rodr{\'{\i}}guez-Fern{\'a}ndez}}]{use04}
{Usero}, A., {Garc{\'{\i}}a-Burillo}, S., {Fuente}, A.,
  {Mart{\'{\i}}n-Pintado}, J., \& {Rodr{\'{\i}}guez-Fern{\'a}ndez}, N.~J. 2004,
  \aap, 419, 897

\bibitem[{{Usero} {et~al.}(2015){Usero}, {Leroy}, {Walter}, {Schruba},
  {Garc{\'{\i}}a-Burillo}, {Sandstrom}, {Bigiel}, {Brinks}, {Kramer},
  {Rosolowsky}, {Schuster}, \& {de Blok}}]{use15}
{Usero}, A., {Leroy}, A.~K., {Walter}, F., {et~al.} 2015, \aj, 150, 115

\bibitem[{{van der Tak} {et~al.}(2007){van der Tak}, {Black}, {Sch{\"o}ier},
  {Jansen}, \& {van Dishoeck}}]{vdt07}
{van der Tak}, F.~F.~S., {Black}, J.~H., {Sch{\"o}ier}, F.~L., {Jansen}, D.~J.,
  \& {van Dishoeck}, E.~F. 2007, \aap, 468, 627

\bibitem[{{van Haarlem} {et~al.}(2013){van Haarlem}, {Wise}, {Gunst}, {Heald},
  {McKean}, {Hessels}, {de Bruyn}, {Nijboer}, {Swinbank}, {Fallows},
  {Brentjens}, {Nelles}, {Beck}, {Falcke}, {Fender}, {H{\"o}randel},
  {Koopmans}, {Mann}, {Miley}, {R{\"o}ttgering}, {Stappers}, {Wijers},
  {Zaroubi}, {van den Akker}, {Alexov}, {Anderson}, {Anderson}, {van Ardenne},
  {Arts}, {Asgekar}, {Avruch}, {Batejat}, {B{\"a}hren}, {Bell}, {Bell}, {van
  Bemmel}, {Bennema}, {Bentum}, {Bernardi}, {Best}, {B{\^i}rzan}, {Bonafede},
  {Boonstra}, {Braun}, {Bregman}, {Breitling}, {van de Brink}, {Broderick},
  {Broekema}, {Brouw}, {Br{\"u}ggen}, {Butcher}, {van Cappellen}, {Ciardi},
  {Coenen}, {Conway}, {Coolen}, {Corstanje}, {Damstra}, {Davies}, {Deller},
  {Dettmar}, {van Diepen}, {Dijkstra}, {Donker}, {Doorduin}, {Dromer}, {Drost},
  {van Duin}, {Eisl{\"o}ffel}, {van Enst}, {Ferrari}, {Frieswijk}, {Gankema},
  {Garrett}, {de Gasperin}, {Gerbers}, {de Geus}, {Grie{\ss}meier}, {Grit},
  {Gruppen}, {Hamaker}, {Hassall}, {Hoeft}, {Holties}, {Horneffer}, {van der
  Horst}, {van Houwelingen}, {Huijgen}, {Iacobelli}, {Intema}, {Jackson},
  {Jelic}, {de Jong}, {Juette}, {Kant}, {Karastergiou}, {Koers}, {Kollen},
  {Kondratiev}, {Kooistra}, {Koopman}, {Koster}, {Kuniyoshi}, {Kramer},
  {Kuper}, {Lambropoulos}, {Law}, {van Leeuwen}, {Lemaitre}, {Loose}, {Maat},
  {Macario}, {Markoff}, {Masters}, {McFadden}, {McKay-Bukowski}, {Meijering},
  {Meulman}, {Mevius}, {Middelberg}, {Millenaar}, {Miller-Jones}, {Mohan},
  {Mol}, {Morawietz}, {Morganti}, {Mulcahy}, {Mulder}, {Munk}, {Nieuwenhuis},
  {van Nieuwpoort}, {Noordam}, {Norden}, {Noutsos}, {Offringa}, {Olofsson},
  {Omar}, {Orr{\'u}}, {Overeem}, {Paas}, {Pandey-Pommier}, {Pandey}, {Pizzo},
  {Polatidis}, {Rafferty}, {Rawlings}, {Reich}, {de Reijer}, {Reitsma},
  {Renting}, {Riemers}, {Rol}, {Romein}, {Roosjen}, {Ruiter}, {Scaife}, {van
  der Schaaf}, {Scheers}, {Schellart}, {Schoenmakers}, {Schoonderbeek},
  {Serylak}, {Shulevski}, {Sluman}, {Smirnov}, {Sobey}, {Spreeuw}, {Steinmetz},
  {Sterks}, {Stiepel}, {Stuurwold}, {Tagger}, {Tang}, {Tasse}, {Thomas},
  {Thoudam}, {Toribio}, {van der Tol}, {Usov}, {van Veelen}, {van der Veen},
  {ter Veen}, {Verbiest}, {Vermeulen}, {Vermaas}, {Vocks}, {Vogt}, {de Vos},
  {van der Wal}, {van Weeren}, {Weggemans}, {Weltevrede}, {White}, {Wijnholds},
  {Wilhelmsson}, {Wucknitz}, {Yatawatta}, {Zarka}, {Zensus}, \& {van
  Zwieten}}]{vanH13}
{van Haarlem}, M.~P., {Wise}, M.~W., {Gunst}, A.~W., {et~al.} 2013, \aap, 556,
  A2

\bibitem[{{Weistrop} {et~al.}(2004){Weistrop}, {Eggers}, {Hancock}, {Nelson},
  {Bachilla}, \& {Kaiser}}]{wei04}
{Weistrop}, D., {Eggers}, D., {Hancock}, M., {et~al.} 2004, \aj, 127, 1360

\bibitem[{{Whitmore}(2007)}]{whi07}
{Whitmore}, B.~C. 2007, in IAU Symposium, Vol. 237, Triggered Star Formation in
  a Turbulent ISM, ed. B.~G. {Elmegreen} \& J.~{Palous}, 222--229

\bibitem[{{Wynn-Williams} \& {Becklin}(1993)}]{wyn93}
{Wynn-Williams}, C.~G. \& {Becklin}, E.~E. 1993, \apj, 412, 535

\end{thebibliography}

\begin{appendix}

\section{WideX spectrum of the NOEMA observations} \label{sec:appendix}

\begin{figure*}[h]
  \centering
    \includegraphics[width=\textwidth]{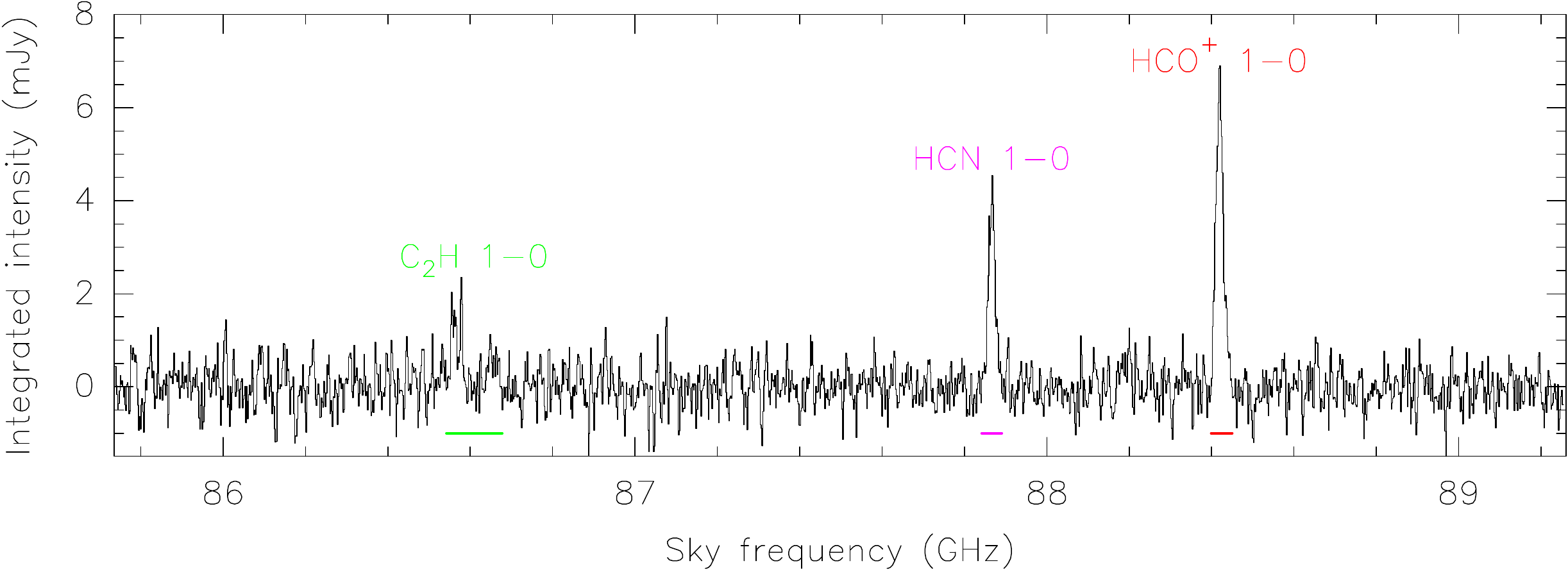}
  \caption{\footnotesize NOEMA spectrum of the WideX 3.6~GHz bandwidth (at $\sim$1.95~MHz ($\sim$~6.6km\,s$^{\rm -1}$) channel spacing) extracted 
   from the naturally weighted data cube. The position of the HCO$^{\rm +}$, HCN and C$_{\rm 2}$H\,1$-$0 lines are indicated in colour. Note that due to 
   hyperfine splitting, C$_{\rm 2}$H\,1$-$0 shows two emission line components.}
  \label{fig:3mm_spectrum}
\end{figure*}

\section{Comparison of the 3~mm and 1~mm continuum images at similar resolution} \label{sec:appendix_cont}

\begin{figure*}[h]
 \centering
  \begin{minipage}[hbt]{0.4975\textwidth}
  \centering
    \includegraphics[width=0.75\textwidth]{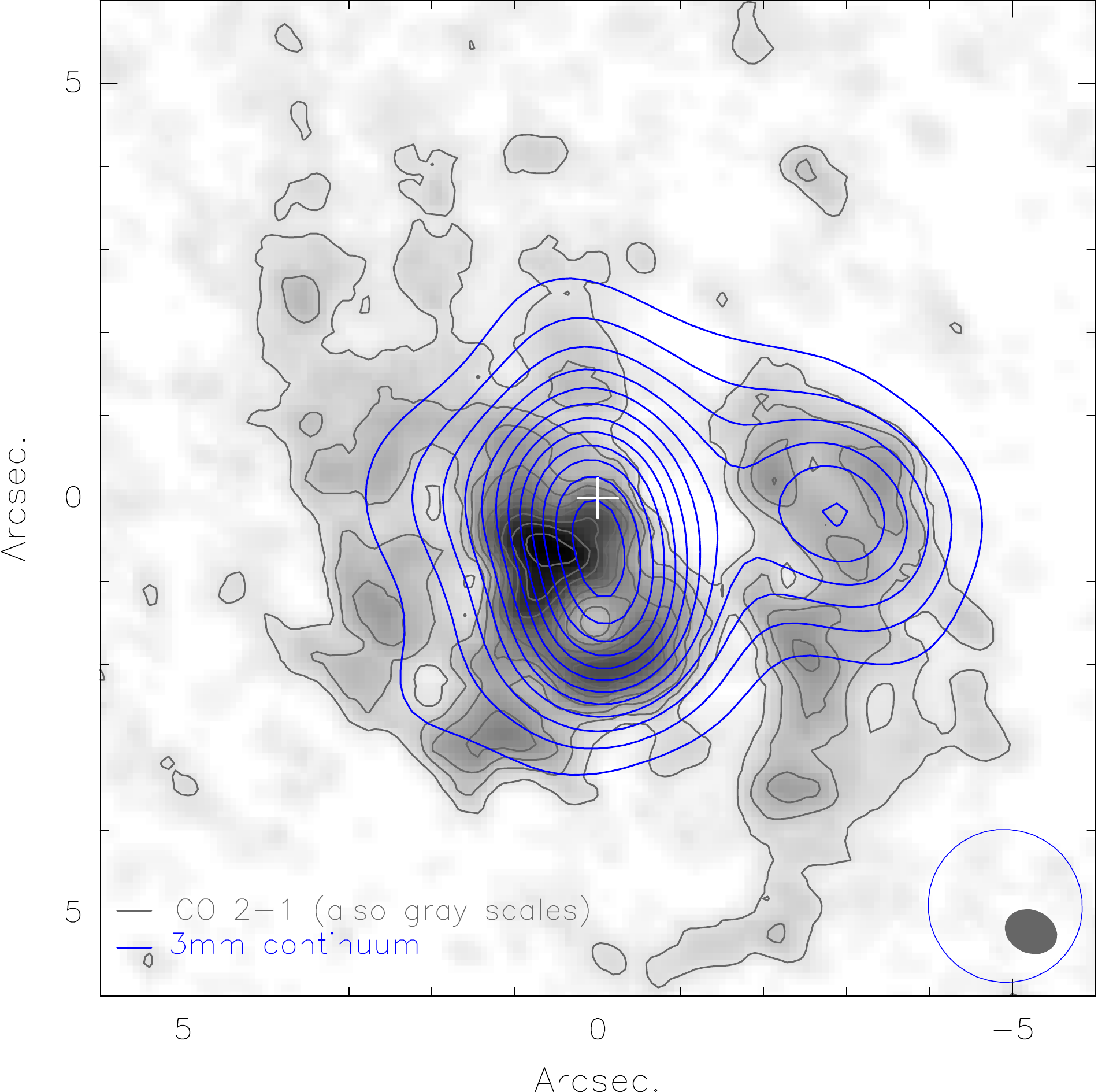}
  \end{minipage}
  \begin{minipage}[hbt]{0.4975\textwidth}
  \centering
    \includegraphics[width=0.75\textwidth]{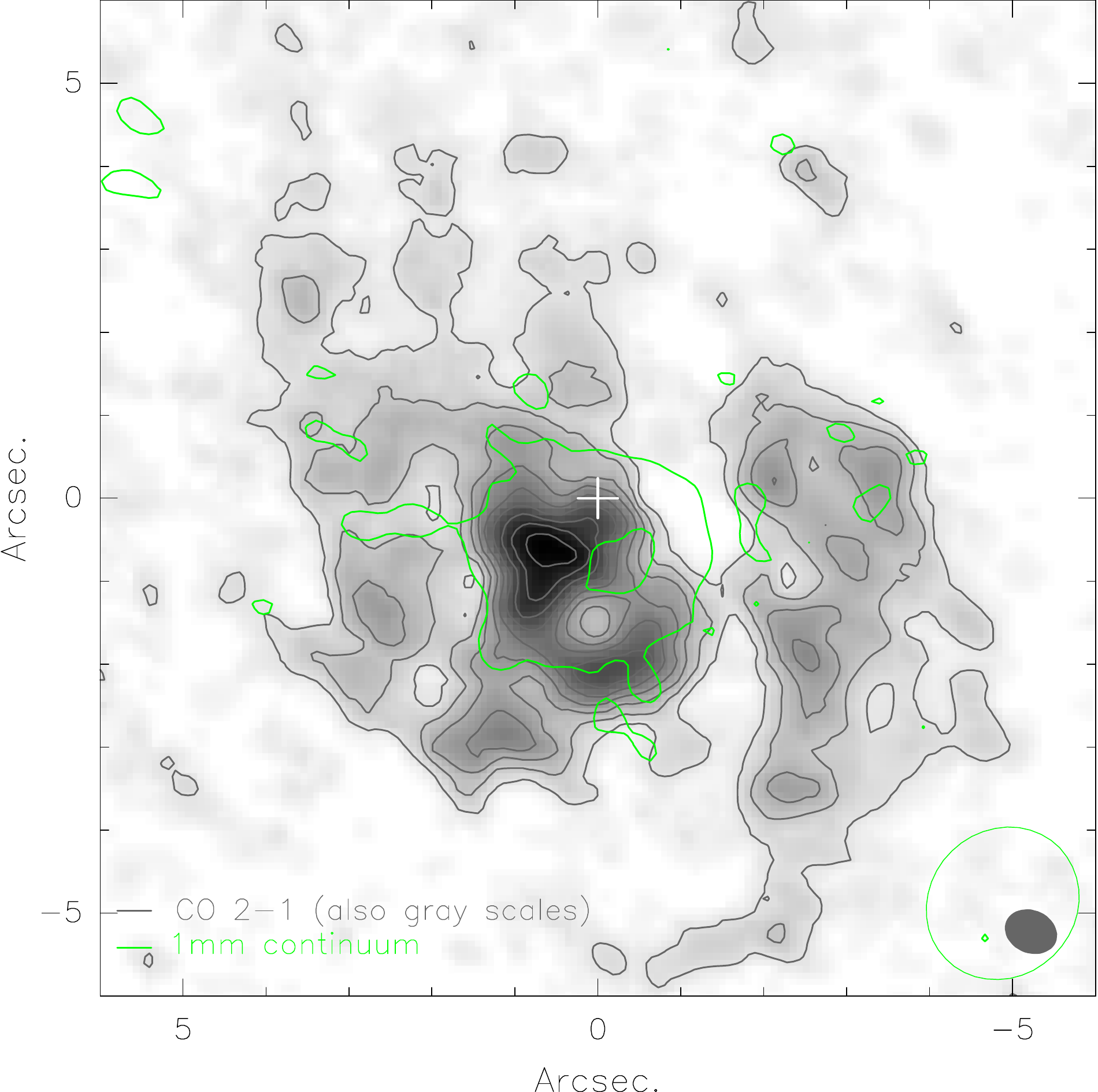}
  \end{minipage}
  \caption{\footnotesize 3~mm (\textit{left}, blue contours) and 1~mm continuum emission (\textit{right}, green contours) at similar spatial 
   resolution on top of the CO\,2$-$1 emission (grey contours and background). The 3~mm continuum contours start at 5$\sigma$ and are spaced in steps of
   5$\sigma$ (1$\sigma$ $\sim$31~$\mu$Jy\,beam$^{\rm -1}$). Contours for the 1~mm continuum emission are at 3 and 6$\sigma$ (1$\sigma$ 
   $\sim$0.5~mJy\,beam$^{\rm -1}$). The main continuum emission peak is located south of the AGN position, close to the center of ``the Eye''. However, 
   the 3~mm continuum peak emission appears elongated; the 1~mm continuum is of more spherical shape. Also, whereas the 3~mm continuum shows a secondary 
   emission peak in the western arm, the 1~mm continuum does not. North is up, east to the left. The beam sizes are 1.9\arcsec\,$\times$\,1.8\arcsec\ 
   for both images.}
  \label{fig:app_cont}
\end{figure*}

\end{appendix}

\end{document}